\documentstyle[epsf,12pt,plen,defphys,defdxm]{article}

\def\thepage{-- \arabic{page} --}

\catcode`\@=11

\long\def\@makecaption#1#2{
   \vskip 10pt 
   \begingroup \small
     \begin{quote}%
       {\bf #1:} #2%
     \end{quote}%
   \endgroup
}
\catcode`\@=12

\begin{document}
\mbox{\ }\\\mbox{\ }\\\mbox{\ }\\

\begin{center}
 \bf   
 THERMODYNAMICS OF THE DOUBLE EXCHANGE SYSTEMS
\end{center}

\mbox{\ }\\\mbox{\ }

\hspace{2.5in}Nobuo Furukawa\\
\mbox{\ }

\hspace{2.5in}\parbox{5in}{
Dept. of Physics, Aoyama Gakuin University,\\
Setagaya, Tokyo 157-8572, Japan
\mbox{\ }\\
\mbox{\ }\\
}

\begin{abstract}

This article gives a comprehensive review on the
recent studies of the double exchange systems
using non-perturbative approaches; the dynamical mean-field
theory and the Monte Carlo method.
Investigations beyond mean-field type treatments
are described.
Taking into account strong spin fluctuations 
which create large changes in conduction electron structure,
finite temperature properties
as well as dynamics of the system are 
calculated.
Comparisons with experimental data 
for colossal magnetoresistance manganites are made.
We show that high Curie temperature ($T_{\rm c}$)
compounds, {\em e.g.} (La,Sr)MnO$_3$,
are canonical double-exchange systems.
Properties of other compounds with lower $T_{\rm c}$ are discussed 
in relation to inhomogeneities of the system including
the issue of phase separation.

\end{abstract}

\section{INTRODUCTION}

The concept of the double-exchange (DE) interaction was
introduced by Zener\cite{Zener51}
in order to explain the ferromagnetism of the
perovskite manganites $A$MnO$_3$.
He considered a Kondo-lattice type model
\begin{equation}
  \Ham = 
  - \sum_{ij,\sigma} t_{ij}
        \left(  c_{i\sigma}^\dagger c_{j\sigma} + h.c. \right)
    -\JH \sum_i \vec\sigma_i \cdot \vec S_i,
    \label{defDEHam}
\end{equation}
where $t$ and $\JH >0$ are $e_{\rm g}$ electron's hopping and 
(ferromagnetic) Hund's
coupling between $e_{\rm g}$ and $t_{2\rm g}$ electrons, respectively.
An effective Hamiltonian in the limit $\JH\to\infty$ 
was introduced by Anderson and Hasegawa\cite{Anderson55}
in the form
\begin{equation}
  \Ham = 
  - \sum_{ij} t(\vec S_i, \vec S_j)
        \left(  \tilde c_{i}^\dagger \tilde c_{j} + h.c. \right).
    \label{defAHHam}
\end{equation}
Mean-field type arguments of these models, including those by
de Gennes,\cite{deGennes60} helped us to discuss the
magnetism of manganite compounds. 

However, recent reinvestigations of these models which are
motivated by the observation of 
colossal magnetoresistance (CMR) phenomena\cite{Chahara93,Jin94}
 in manganites
revealed that such simplified treatments are insufficient to
discuss quantitative nature of the models. 
Let us show an example. Curie temperature $T_{\rm c}$ of the model 
has been estimated by Millis {\em et al.}\cite{Millis95}
 using the mean-field
type discussion. Using some appropriate values for $t$ and $\JH$,
they found that the mean-field $T_{\rm c}$ of the model
is much larger than those for perovskite manganites
(La,Sr)MnO$_3$, and concluded that double-exchange alone
is insufficient to explain the thermodynamics of these manganites.
However, as we will show in this article, accurate treatments for the
model give suppression of $T_{\rm c}$ from the mean-field value,
and show consistency with (La,Sr)MnO$_3$ data.

Another important point which will be discussed here is
the systematic understanding of experiments.
For example, it is known that CMR manganites  exhibit metal-insulator
transition at Curie temperature. This statement is, however,
inaccurate. 
By varying compositions, varieties of phases with different
properties have been known.\cite{Ramirez97}
A typical reference compound La$_{2/3}$Sr$_{1/3}$MnO$_3$
shows no metal-insulator transition. The resistivity 
always increases by the increase of
temperature, {\em i.e.} ${\rm d}\rho(T)/{\rm d}T >0$
even above $T_{\rm c}$. Its absolute value in the
paramagnetic phase is around the Mott's limit value,
and may be explained by the DE model.
On the other hand, La$_{2/3}$Ca$_{1/3}$MnO$_3$
which is often reffered to as an ``optimal'' CMR material
shows insulating behavior above $T_{\rm c}$ with
much larger $\rho(T)$.
Thus we have to be mindful to make distinctions between
various compositions of manganites in a systematic manner.

The purpose of this article is to solve such misunderstandings
and confusions in both theoretical and experimental studies.
We review
the investigation of the DE model by
the author.\cite{Furukawa94,Furukawa95b,Furukawa95c,Furukawa95d,Furukawa96,Furukawa97,Furukawa98,Furukawa98x,Yunoki98}
The first part of this article shows the finite temperature behaviors
in the DE systems.
In \S2 we introduce the model and the method (dynamical mean-field
approach and the Monte Carlo calculation).
In \S3 the results for infinitely large spin (classical spin limit)
is presented, and the $1/S$ correction is shown in \S4.
The second part is devoted for discussions with
respect to comparisons of model behavior with experimental data.
Comparison with experiments.
 are shown in \S5.
Section 6 is devoted for concluding remarks.

\section{MODEL AND METHODS}

\subsection{Double Exchange Model in the Large Spin Limit}

\subsubsection{Model.}
The compound LaMnO$_3$ has four 3d electrons per atom
in the $(t_{2\rm g})^3 (e_{\rm g})$ configuration.
Due to Hund's coupling, these electrons have the high spin state,
{\em i.e.} spin parallel configuration. By substitution of the
La site with alkaline-earth divalent ions, holes are doped as 
carriers which is considered to enter the $e_{\rm g}$ orbitals.

As briefly mentioned in the previous section, 
Zener\cite{Zener51} introduced a Kondo-lattice type Hamiltonian
(\ref{defDEHam}) with ferromagnetic spin exchange $\JH$
between localized spins and itinerant electrons.
In manganites ($R$,$A$)MnO$_3$, Hund's coupling $\JH$ is estimated
to be a few eV while electron hopping $t$ be in the order of 0.1eV.
Therefore, we have to deal with a strong coupling region $\JH\gg t$.
In order to discuss the ferromagnetism of the model in the
metallic region, Zener introduced  the notion of the
``double exchange'' interaction.

In some cases, the limit $\JH\to\infty$ 
first studied by Anderson and Hasegawa\cite{Anderson55}
is called
the DE model. However, in order to avoid complications,
we call the model (\ref{defDEHam}) in the strong coupling region
as DE model.
In the weak coupling limit $\JH/t \ll 1$, the model is
often referred to as the $s$-$d$ model and studied
as a model hamiltonian for the magnetic semiconductors.

The classical rotator limit,
or equivalently large spin limit $S=\infty$,
has been introduced by Anderson and Hasegawa.\cite{Anderson55}
In manganites, we consider the case where
 the localized spin is in a high-spin state ($S=3/2$)
with the ferromagnetic coupling,
the effect of quantum exchange
seems to be less relevant compared to thermal fluctuations,
at least in low energy physics.
However,
 the role of quantum exchanges might give
non-trivial effects in the system
in the region with less thermal fluctuations.
Such issues are left for future studies.

In this article,
we introduce a model with finite $\JH$ and infinite $S$,
\bequ
  \Ham_{S=\infty} = 
  - \sum_{ij,\sigma} t_{ij}
        \left(  c_{i\sigma}\dags c_{j\sigma} + h.c. \right)
    -\JH \sum_i \vec \sigma_i \cdot \vec m_i.
    \label{HamSinfty} \label{HamDXM}
\eequ
Hereafter we express the localized (classical) spin  by
 $ \vec m_i = (m_i{}^x, m_i{}^y, m_i{}^z)$ 
with the normalization $|\vec m|^2 = 1$.

\subsubsection{Previous investigations.}

Anderson-Hasegawa\cite{Anderson55} as well as
de Gennes\cite{deGennes60} studied the magnetism of the model
with localized spins treated as static, {\em i.e.} neglecting
spin fluctuations, they made a
search for energetically favored spin configurations
as well as a  mean-field calculation for finite temperatures.
However, if we consider the system at finite temperature,
especially around $T\sim T_{\rm c}$, spin fluctuations $\delta S_i$
become important. Especially, in our case of $\JH \gg t$,
such spin fluctuations give large effects to
electronic structures since $\JH \delta S \gg t$.

The point of interest for us in relationship with CMR phenomena
is the change of resistivity $\rho$ at around $T_{\rm c}$.
Mean-field type treatments are not justified
for this purpose.
An alternative approach has been made by Kubo and Ohata.\cite{Kubo72}
They phenomenologically assumed the electron self-energy $\Sigma$ in the form
\begin{equation}
 \tau^{-1} = \Im \Sigma \propto 1-M^2,
\end{equation}
where $\tau$ is the quasiparticle lifetime,
and calculated the resistivity using the Drude's formula
$\rho \propto \tau^{-1}$.
Within this phenomenological treatment,
the result qualitatively helps us to understand
 the magnetoresistance (MR)
of the DE model via spin disorder scattering mechanism.
However, in a quantitative way, it fails to
reproduce the MR of (La,Sr)MnO$_3$ as we will show in
\S\ref{Experimental}.

\subsubsection{Beyond the previous theories.}

Thus, in order to understand the behaviour of the DE model
and to make direct comparisons with experimental data,
it is very important to treat the model in an accurate way.
The methods have to be an unbiased non-perturbative approaches
since $\JH\gg t$, and must 
take into account 
the effect of spin fluctuations to calculate finite temperature
behaviors.

Here we introduce two methods: The dynamical mean-field (DFM)
theory and the Monte Carlo calculation.
Both of these methods give us the electronic states
at finite temperature including $T\sim T_{\rm c}$.
They are unbiased and become exact in the limit
of large coordination number and large system size, respectively.
One of the advantages of these methods, in the viewpoint
of comparison with experiments, is that it is easy to obtain
dynamical quantities such as density of states (DOS)
and optical conductivity $\sigma(\omega)$.

\subsection{Dynamical mean-field theory}

We introduce the DMF method
for the double-exchange model.
For a general review of the field, see the review articles
in refs.~\citen{Pruschke95} and \citen{Georges96}.
Within the general scheme of the DMF, we treat a lattice system by
considering a single-site coupled with ``electron bath'',
or the time-dependent mean-field $\tilde G_0$.
This method becomes exact in the large coordination number limit
or equivalently large spatial dimension 
limit.\cite{Metzner89,Muller-Hartmann89}

Generic part of the DMF treatment is as follows.
Solving a model-specific single-site problem, we obtain the self-energy
$\tilde \Sigma(\MFreq)$ from $\tilde G_0$.
Lattice Green's function is approximated by
\begin{equation}
  G(k,\MFreq) = [{\MFreq - (\varepsilon -\mu) - \tilde\Sigma(\MFreq)}]^{-1}.
\end{equation}
The local Green's function is defined by
\begin{equation}
 G_{\rm loc}(\MFreq) \equiv \frac1N \sum_k G(k,\MFreq)
 \label{defGlocal}
\end{equation}
Since $\tilde \Sigma$ is $k$-independent,
$k$ dependence of $G(k,\MFreq)$ comes through the energy dispersion
and we have
\begin{equation}
 G_{\rm loc}(\MFreq) =
   \int \rmd \varepsilon 
        {N_0(\varepsilon) }
        [ \MFreq - (\varepsilon -\mu) - \tilde\Sigma(\MFreq)]^{-1}.
                \label{intGlocal}
\end{equation}
 $N_0(\varepsilon)$ is the DOS
for the noninteracting lattice system.
The information of the lattice geometry is
included through the noninteracting DOS in eq.~(\ref{intGlocal}).
The method is applicable to  finite size systems
by taking the sum over $k$ points in eq.~(\ref{defGlocal})
in the discrete $k$-space.
We self-consistently obtain the time-dependent mean-field $\tilde G_0$ as
\begin{equation}
  \tilde G_0(\MFreq) = \left(
        G_{\rm loc}^{-1}(\MFreq) + \tilde\Sigma(\MFreq)
        \right)^{-1}.
        \label{SCEcondition}
\end{equation}

Let us now discuss the model-specific part.\cite{Furukawa94}
For the present system (\ref{HamSinfty}),
the action of the effective single-site model is described as
\begin{eqnarray}
  \tilde S &=&
  - \int_0^\beta \rmd \tau_1
   \int_0^\beta \rmd \tau_2\ 
     \Psi^* (\tau_1) \tilde G_0^{-1}(\tau_1 - \tau_2) \Psi(\tau_2)
        \nonumber \\
  & &  -\JH  \int_0^\beta \rmd \tau \ 
      \vecn \cdot \Psi^*(\tau) 
           \vec \sigma  \Psi(\tau)  .
        \label{Action}
\end{eqnarray}
Here $\Psi^* = ( c_\uparrow^*, c_\downarrow^*)$ is
the Grassmann variables in the spinor notation.
Green's function in the imaginary time is calculated as
\begin{eqnarray}
  \tilde G(\MFreq) &= &
  \braket{ \left(\tilde G_0^{-1}(\MFreq) +
            \JH \vecn \cdot \vec \sigma \right)^{-1}}
   \nonumber\\
  &=&
   \frac1{\tilde Z} \int \rmd\Omega P(\vecn) 
        \left(  \tilde G_0^{-1}(\MFreq) +
            \JH \vecn \cdot \vec \sigma \right)^{-1} .
        \label{GreensFunctionImp}
\end{eqnarray}
$P(\vecn)$ is the Boltzmann factor for the spin 
\begin{equation}
   P(\vecn) = \frac1{\tilde Z}  \exp[-\tilde S_{\rm eff}(\vecn)] ,
        \label{defBoltzmann}
\end{equation}
where $\tilde S_{\rm eff}$ is the
effective action for the spin
\begin{eqnarray}
  &&\tilde S_{\rm eff}(\vecn)
  =  -\log {\rm Tr_{\rm F}} \exp(-\tilde S) 
        \nonumber \\
  &&\ =  - \sum_n \log \det \left[ 
       \frac1{\MFreq} (\tilde G_0^{-1}(\rmi \omega_n) + \JH\vecn\vec\sigma)
    \right] \rme^{\MFreq 0_{+}} .\quad
        \label{defAction}
\end{eqnarray}
$\tilde Z$ is the partition function
\begin{eqnarray}
  \tilde Z 
   &=&  \int \rmd\Omega_{\vecn} 
  \int {\cal D}\Psi^* {\cal D}\Psi \exp(-\tilde S) 
  \nonumber\\
  &=& \int \rmd\Omega_{\vecn} 
        \exp[-\tilde S_{\rm eff}(\vecn)].
      \label{PartitionFunction2}
\end{eqnarray}
The self energy for the single-site system $\tilde\Sigma$ is obtained from
$   \tilde \Sigma(\MFreq) = \tilde G_0 ^{-1}(\MFreq) - \tilde G ^{-1}(\MFreq)$.

Magnetization of the local spin is obtained by
\bequ
\braket{\vecn}
  =     \int \rmd\Omega_{\vecn} 
        P(\vecn) \vecn.
\eequ
Hereafter we take the axis of the magnetization in $z$ direction and
the order parameter is expressed as $M = \braket{m_z}$.
Transport properties are obtained through the
Kubo formula.
Conductivity in $D=\infty$ is  
calculated as\cite{Moller92,Pruschke93,Pruschke93a}
\beqarr
  \sigma(\omega) &=& \sigma_0
  \sum_\sigma \int \rmd \omega' \ I_\sigma(\omega',\omega'+\omega)
   \frac{ f(\omega') - f(\omega'+\omega)}{\omega},
   \label{Optcond}
\eeqarr
where
\bequ
   I_\sigma(\omega_1,\omega_2) = \int N_0(\epsilon) \rmd \epsilon \
     W^2 A_\sigma(\epsilon,\omega_1) A_\sigma(\epsilon,\omega_2).
\eequ
Here, the spectral weight function is defined by
\bequ
  A_\sigma(\epsilon,\omega) = 
  -\frac1\pi {\rm Im} G_\sigma(\epsilon,\omega+\rmi\eta) ,
\eequ
while $f$ is the Fermi distribution function.
The constant $\sigma_0$ gives the unit of conductivity.
In this formula, we used that the vertex correction cancels in the
 conductivity calculation at infinite-dimensional limit.\cite{Khurana90}
Thermopower is also calculated: The Seebeck coefficient is
obtained as\cite{Schweitzer91,Pruschke95}
\bequ
   S = \frac{1}{eT}  \frac{L_2}{L_1},
\eequ
where $L_k$ ($k=1,2$) is defined by
\bequ
  L_k = \sum_\sigma \int \rmd \omega 
        \left( -\frac{\partial f(\omega)}{\partial \omega} \right)
        I_\sigma(\omega,\omega) (\beta\omega)^{k-1}.
\eequ

The method is easily expanded to a
 Bethe lattice with two-sublattice symmetry.
In this case, 
  magnetic phases
with ferromagnetic and antiferromagnetic  order parameters
can be considered simultaneously.
We introduce $\alpha=A,B$ sublattice indices for
 the Weiss fields  $\tilde G_{0\alpha}(\MFreq) $,
and solve coupled self-consistency equations.
The formula to calculate the Green's function 
 is now given by
\bequ
  \tilde G_\alpha(\MFreq) =
  \int \rmd \Omega_\alpha P_\alpha(\vecn) 
       \left(  \tilde G_{0\alpha}^{-1}(\MFreq) +
            J \vecn \cdot \vec \sigma \right)^{-1}.
        \label{defGfunalpha}
\eequ
Here, the Boltzmann weight for the configuration of local spin $P_\alpha(\vecn)
$
is calculated from the effective action $\tilde S_{\alpha}$, 
\begin{eqnarray}
  \tilde S_{\alpha}(\vecn) &= &
  -\log {\rm Tr_{\rm F}} \exp[-\tilde S(\tilde G_{0\alpha},\vecn)] ,
                 \\
  P_\alpha(\vecn) &=&
          \exp[-\tilde S_{\alpha}(\vecn)] / {\tilde Z_\alpha} ,
                 \\
  \tilde Z_\alpha &=&
          \int \rmd \Omega_\alpha \exp[-\tilde S_{\alpha}(\vecn)].
\end{eqnarray}
Integration over
DOS as in eq.~(\ref{SCEcondition}) gives the self-consistent
mapping relation\cite{Rozenberg94}
\beqarr
   \tilde G_{0A}{}^{-1}(\MFreq) &=&
                 \MFreq + \mu - \tilde G_B ( \MFreq) \  W^2/4 ,
                \nonumber\\
   \tilde G_{0B}{}^{-1}(\MFreq) &=&
                 \MFreq + \mu -\tilde  G_A ( \MFreq) \ W^2/4 .
        \label{defSCCG}
\eeqarr
Now the self-consistency equations
 (\ref{defGfunalpha})-(\ref{defSCCG}) form a closed set.
Within this approach, we can study the instability and the formation
of magnetic ordering with ferromagnetic, antiferromagnetic, and
canted antiferromagnetic symmetries.


\subsection{Analytical solutions of the dynamical mean-field theory}

\label{DMFanalytical}

In some limiting cases, analytical solutions are available
for the DMF calculations.
Analytical expressions are in general quite useful to obtain intuitions.

\subsubsection{Paramagnetic phase.}
For the paramagnetic solution, Green's function is a
scalar function with respect to spin rotation, so we have
$  \tilde G_0(\MFreq) =  \tilde g_0(\MFreq) \eignmat$ where
$\eignmat$ is the $2\times2$ eigenmatrix.
Then, we have
\begin{equation}
  \tilde G(\MFreq) = 
    \braket{
      \frac{ 
            \tilde g_0(\MFreq)^{-1} \eignmat - \JH \vecn \vec \sigma
      }
      {  \tilde g_0(\MFreq)^{-2} - \JH{}^2 |\vecn|^2 }
    }.
   \label{GFunAve}
\end{equation}                  
Since $\braket{\vec m} = 0$ and $\braket{|\vec m|^2} = 1$, we have
\begin{equation}
   \tilde G(\MFreq) 
     =  \frac{  \tilde g_0(\MFreq)^{-1} }
             {  \tilde g_0(\MFreq)^{-2} - \JH{}^2}
      \eignmat 
     = \frac12 \left(
                  \frac1{  \tilde g_0(\MFreq)^{-1} + \JH}
                + \frac1{  \tilde g_0(\MFreq)^{-1} + \JH}
                \right) \eignmat .
     \label{GFunSolution}
\end{equation}
The self-energy is then given by 
\bequ
  \Sigma(\MFreq) = 
    \tilde G_0^{-1}(\MFreq) - \tilde G^{-1}(\MFreq) 
   = \JH{}^2 \tilde G_0(\MFreq).
    \label{SelfEnergy}
\eequ

From the derivation, we see that the Green's function 
(\ref{GFunSolution}) is the same as that of
the system with the Ising substrate spin
$\vecn = (0,0,\pm1)$.
Furthermore, 
the Green's function (\ref{GFunSolution}) shares the same
analytical structure as that of the 
infinite-dimensional Falicov-Kimball model (FKM) (or, simplified
Hubbard model)
\bequ
        \Ham_{\rm FKM} = -\sum_{ij} t_{ij} (c_i\dags c_j + h.c.)
    + U \sum_i c_i\dags c_i f_i\dags f_i
\eequ
at $\brakets{n_{\rm f}}=1/2$.
Green's function of the FKM in infinite dimension
is described as\cite{Brandt89,Si92}
\bequ
  \tilde G_{\rm FKM}(\MFreq) 
    =    \frac{1-\brakets{n_{\rm f}}}{ \tilde G_0^{-1}(\MFreq)}
      +  \frac{\brakets{n_{\rm f}}}{ \tilde G_0^{-1}(\MFreq) - U}.
        \label{GFunFK}
\eequ
Then, the  Green's functions (\ref{GFunSolution}) and (\ref{GFunFK})
share the same analytical structure at $\brakets{n_{\rm f}}=1/2$.
In the FKM, the c-electrons are scattered by the
charge fluctuations of the localized
f-electrons,
which corresponds to the
scattering process of the itinerant electrons by the localized
spins in the DE model.

Hence, thermodynamical properties
of the DE model in $D=\infty$ and $S=\infty$
can be understood from the nature of
the FKM in $D=\infty$ which has been studied intensively.
For example,
the imaginary part of the self-energy at the fermi level is finite 
$  \Im \Sigma(0) \ne 0$ in the paramagnetic phase.\cite{Moller92,Si92}

\begin{figure}
\epsfxsize=15cm\epsfbox{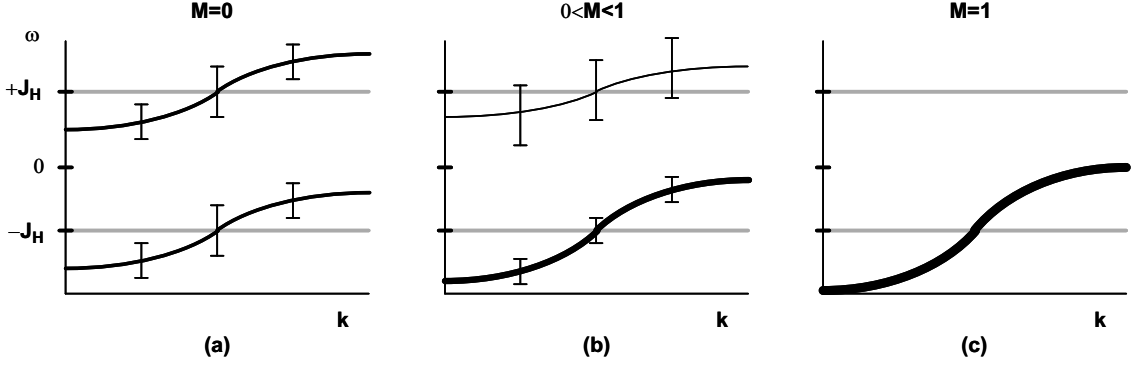}
\caption[]{Schematic behavior in the spectral function for 
up-spin electrons,
at (a) paramagnetic state $M=0$, (b) ferromagnetic state
at $0<T<T_{\rm c}$, and (c) at the ground state $T=0$ and $M=1$.
Solid curves illustrate peak positions of $A(k,\omega)$.
Width of the curves represent height of the peak 
(quasiparticle weight), while
error bars
represent the linewidth (inverse of lifetime).
 Grey lines are guides to eyes.
Spectral functions for down-spin electrons are obtained by exchanging
upper part ($\omega\sim\JH$) and lower part ($\omega\sim -\JH$).
Also see Fig.~\ref{FigAw} in \S\ref{DOSstructure} for actual data.
}
\label{FigAw-G}
\end{figure}

In Fig.~\ref{FigAw-G}(a) we schematically illustrate
the spectral function $A(k,\omega)$ in the paramagnetic phase.
As in the case for FKM, the spectral weight is split into
two parts at $\omega \sim \pm \JH$
for sufficiently large Hund's coupling $\JH \gg W$.
For the semi-circular density of states with the bandwidth $W$,
we have the metal-insulator transition\cite{vanDongen92} at 
$ J_{\rm H}{}^{\rm c} = 0.5 W$,
which has the  Hubbard-III like nature.
Kondo resonance peak, which is seen in the 
Hubbard model\cite{Georges92,Jarrel93,Sakai94}
is missing in this model, since the quantum exchange process is absent.
As the magnetic moment is induced, the imaginary part of the
self-energy decreases
because the thermal fluctuation of spins decreases.
At the ground state, spins are magnetically ordered.
For the ferromagnetic ground state, there exists
two free electron bands which are energetically
split by $ \JH$ exchange interactions.

\subsubsection{Infinite $\JH$ limit --Green's function--.}
In the case of Lorentzian DOS, Green's function is
easily obtained in the limit $\JH\to\infty$.\cite{Furukawa95c}
We consider the hole doped region where $\mu\sim -\JH$.
In the case of the Lorentzian DOS, the self-consistency
equation gives
\beqarr
  G_0 (\omega+\rmi\eta) 
  &=& (\Omega - \JH + \rmi  W)^{-1}.
    \label{defG0}
\eeqarr
Here, chemical potential is 
$ \mu = -\JH + \delta\mu$ where $ \delta\mu = O(W)$,
and $\Omega \equiv \omega + \delta\mu = O(W)$ 
is the energy which is measured
from the center of the lower sub-band $-\JH$.

Magnetic field in the $z$ direction 
is applied to the localized spins in the paramagnetic phase,
and  the induced magnetization is expressed as $ M = \braket{m_z}$.
Since $G_0$ in eq.~(\ref{defG0}) is spin independent even in the
spin polarized cases,
eqs.~(\ref{GFunAve}) and (\ref{defG0}) gives
\beqarr
  G_\sigma(\omega+\rmi\eta) 
   &=& \frac{ (\Omega - \JH + \rmi W) - \JH M\sigma}
            { (\Omega - \JH + \rmi W)^2 - \JH{}^2}
                \nonumber \\
   &=&  \frac{1+M\sigma}2\frac1{\Omega + \rmi W} + O(1/\JH).
    \label{defG}
\eeqarr
At $\JH/W\to\infty$, the spectral weight is calculated as
\beqarr
  A_\sigma(\omega) &=& -\frac1\pi \Im G_\sigma(\omega+\rmi\eta)
  = \frac{1+M\sigma}{2}\cdot\frac1{\pi}\frac{W}{\Omega^2+W^2}.
  \label{defQPdos}
\eeqarr
We see that the center of the spectral weight
is indeed shifted to $-\JH$. The amplitude of $A_\sigma$ is proportional to the
population of the local spins parallel to $\sigma$, which indicates
that the electronic states
 that are anti-parallel to the local spin are projected out.
The self-energy is calculated from eqs.~(\ref{defG0})
and (\ref{defG}) as
\beqarr
  \Sigma_\sigma(\omega+\rmi\eta) 
  &=& -\JH - \frac{1-M\sigma}{1+M\sigma}(\Omega+\rmi W).
  \label{defSigma}
\eeqarr
Eq.~(\ref{defSigma}) gives
 $\Re \Sigma \sim -\JH$, so the shift in $\mu$ is
self-consistently justified again. 

Similarly, Green's function at $\omega' \sim 2\JH$, namely at
the upper subband, is described as follows.
Using $\omega'= \omega+ 2\JH$, where $\omega=O(W)$,
Green's function is given by
\begin{equation}
  G_\sigma(\omega+2\JH+\rmi\eta) 
   =  \frac{1-M\sigma}2\frac1{\Omega + \rmi W} + O(1/\JH),
    \label{defGUP}
\end{equation}
and the spectral weight is calculated as
\bequ
  A_\sigma(\omega+2\JH) 
  = \frac{1-M\sigma}{2}\cdot\frac1{\pi}\frac{W}{\Omega^2+W^2}.
  \label{defQPdosUP}
\eequ
From eqs.~(\ref{defQPdos}) and (\ref{defQPdosUP}), we see the transfer of the
spectral weight by magnetization.

From above equations,  Green's function is given in the form
\begin{eqnarray}
  G_\sigma(k,\omega;M) 
   &=& \phantom{+} \frac{z_\sigma^{\rm (l)}(M)}
      { \omega + \JH +\mu 
         - \zeta_{\sigma}^{\rm (l)}(k;M) + \rmi \Gamma_\sigma^{\rm (l)}(M)}
         \nonumber\\
 &&  + \frac{z_\sigma^{\rm (u)}(M)}
      { \omega - \JH + \mu 
         - \zeta_{\sigma}^{\rm (u)}(k;M) + \rmi \Gamma_\sigma^{\rm (u)}(M)},
         \label{defGall}
\end{eqnarray}
asymptotically at $\JH/W\to \infty$.
In this limit, 
Green's function is a sum of contributions from lower subbandrs (l)
and upper subband (u).
Quasiparticle residue is given by
\begin{equation}
 z_\sigma^{\rm (l)}(M) = P_\sigma^{\rm +}(M), \qquad
 z_\sigma^{\rm (u)}(M) = P_\sigma^{\rm -}(M),
\end{equation}
quasiparticle dispersion relation is described as
\begin{equation}
 \zeta_{\sigma}^{\rm (l)}(k;M) = P_\sigma^{\rm +}(M) \varepsilon_k,
    \qquad
 \zeta_{\sigma}^{\rm (u)}(k;M) = P_\sigma^{\rm -}(M) \varepsilon_k,
\end{equation} 
and the quasiparticle linewidth is in the form
\begin{equation}
  \Gamma_\sigma^{\rm (l)}(M) = P_\sigma^{\rm -}(M) W, 
   \qquad
  \Gamma_\sigma^{\rm (u)}(M) = P_\sigma^{\rm +}(M) W.
\end{equation}
Here, $P$ are the function of spin polarization 
\begin{equation}
  P_\sigma^{\rm + }(M) \equiv \frac{1+M\sigma}{2},\qquad
  P_\sigma^{\rm - }(M) \equiv \frac{1-M\sigma}{2}.\qquad
\end{equation}

In Fig.~\ref{FigAw-G} we schematically show the
spectral function for the up-spin electron
 $A_\uparrow(k,\omega)=-\Im G_\uparrow(k,\omega)/\pi$
calculated from eq.~(\ref{defGall}).
For the down-spin electrons, $A_\downarrow(k,\omega)$ is
obtained by replacing upper and lower subbands,
{\em e.g.} $z_\downarrow^{\rm (l)}= z_\uparrow^{\rm (u)}$, etc.
In the paramagnetic phase (Fig.~\ref{FigAw-G}(a)),
quasiparticle dispersion are split into lower and upper subbands
at $\omega\sim \pm\JH$
with quasiparticle weight $z_\uparrow^{\rm (l)}=z_\uparrow^{\rm (u)}=1/2$.
As magnetization is increased below $T_{\rm c}$, the lower subband
gains quasiparticle weight $z_\uparrow^{\rm (l)} = (1+M)/2$
and that for upper subband decreases as $z_\uparrow^{\rm (u)} = (1-M)/2$.
At the ground state with perfect spin polarization $M=1$,
the electronic Hamiltonian describes a free electron system under
Zeeman splitting field. Hence there exists only lower subband
for the up spin electron.
This limit is described as $z_\uparrow^{\rm (l)} = 1$ and
 $z_\uparrow^{\rm (u)} = 0$. 

In the paramagnetic phase $M=0$ we see
 $\Im \Sigma = -W$,
which means that the quasi-particle excitation is very incoherent;
the lifetime of a  quasi-particle is comparable with the
time scale that an electron transfers from site to site.
This result justifies us to take the $D=\infty$ limit which is
essentially a single-site treatment.

In the following section (\S\ref{DOSstructure}),
we will discuss the behavior of the spectral function
for general cases.

\subsubsection{Infinite $\JH$ limit --Curie temperature--.}
In the semicircular DOS case,
 we consider the ferromagnetic state under doping at $\JH \gg W$.
We set
\bequ
  G_{0}{}^{-1}(\MFreq) = (\MFreq + R_n) \hat I + 
        Q_n \hat\sigma_z
\eequ
and $\mu = -\JH + \delta\mu$.
In order to keep the carrier concentration finite, we take
 the limit $\JH\to\infty$ with keeping $\delta\mu = O(W)$. In this limit,
we have
\bequ
  R_n = -\frac{W^2}{8}
        \braket{\frac1{ z_n + R_n + m_z Q_n} } ,\qquad
  Q_n =
 -\frac{W^2}{8}
        \braket{\frac{m_z}{ z_n + R_n + m_z Q_n} } ,
\eequ
where $z_n = \MFreq + \delta\mu$.
The Boltzmann weight is calculated from eq.~(\ref{defAction}) as
\bequ
  P(\vec m) \propto \exp \left[
        \sum_n \log\left( 1 + \frac{ R_n + Q_n m_z}{z_n} \right)
                \right].
\label{defBoltzDE}
\eequ
This equation tells us that the model 
is not simply mapped to the Heisenberg model.
The  Boltzmann weight of the Heisenberg model in infinite dimension
 is expressed as
$   P_{\rm Heis}(\vec m) \sim \exp (-\beta h_{\rm eff} m_z)$,
which contradicts with that of the DE model,
{\em i.e.} $ P(\vec m) \ne P_{\rm Heis}$.
This reflects the fact that the itinerant ferromagnet has
smaller $T_{\rm c}$ due to the spin fluctuation affecting
the itinerant electrons, compared to the insulating ferromagnet
with the same spin stiffness.

\begin{figure}
\epsfxsize=8cm\epsfbox{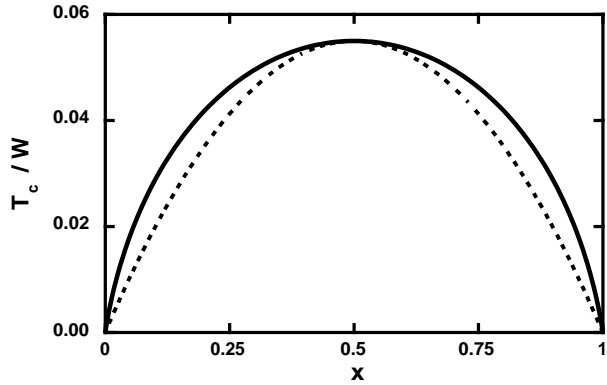}
\caption[]{Curie temperature for the semicircular DOS at $\JH=\infty$
 (solid curve). Dotted curve is the form $T_{\rm c} \propto x(1-x)$.
}

\label{FigTcJHinfty}
\end{figure}

Let us calculate $T_{\rm c}$. At $T \sim T_{\rm c}$, 
we have $M=\braket{m} \ll 1$ and
\beqarr
  R_n &=&  -\frac{W^2}{8}
        \frac1{ z_n + R_n }  + O(M),
     \label{solveRn}  \\
  Q_n &=&
  -\frac{W^2}{8}
        \braket{\frac{m_z}{ z_n + R_n } - \frac{m_z^2}{( z_n + R_n)^2 }Q_n}
         + O(M^2).
     \label{solveQn}  
\eeqarr
By solving eqs.~(\ref{solveRn}) and (\ref{solveQn}),  we have
\begin{equation}
  R_n = \sqrt{z_n{}^2 - W^2/4} - z_n,\qquad
  \frac{Q_n}{M} = \frac{R_n}{1 - 8 \braket{m_z{}^2} R_n^2 / W^2}.
   \label{solutionRQ}
\eequ
Here, $\braket{m_z{}^2} \equiv \int \rmd \Omega m_z{}^2 = 1/3$ at $S=\infty$.
The Boltzmann weight is given by
\bequ
  P(\vec m)\propto \exp\left[ \sum_n \frac{Q_n}{z_n + R_n} m_z \right]
  = \exp( -\beta J_{\rm eff} M m_z)
\eequ
where
\bequ
  J_{\rm eff}(\beta) = \frac1{\beta} \sum_n \frac{ 8R_n^2/W^2}
                        {1 - 8 R_n^2/(3W^2)}.
        \label{JeffFerro}
\eequ
Then, the partition function is identical to that of the Heisenberg model
with exchange coupling $J_{\rm eff}$, and $T_{\rm c}$ is obtained from
\begin{equation}
  T_{\rm c} = \left.
       \frac1{\beta} \sum_n \frac{ 8R_n^2}
        {3W^2 - 8 R_n^2}
              \right|_{\beta=1/T_{\rm c}}.
  \label{SCETC}
\end{equation}

Solving eqs.~(\ref{solutionRQ}) and (\ref{SCETC}) self-consistently,
we obtain $T_{\rm c}$ as a function of $\mu$, while the carrier number
is calculated directly from Green's function $G(\rmi \omega_n)$.
In Fig.~\ref{FigTcJHinfty} we plot $T_{\rm c}$ as a function of doping $x$.
We see that $T_{\rm c}$ has a maximum at $x=0.5$ with $T_{\rm c} \sim 0.05W$.
Then, using $\beta W \ll 1$ 
we may approximate the summation in eq.~(\ref{SCETC}) by integration,
\begin{equation}
    T_{\rm c} = \frac1{2\pi} \int_{-\infty}^\infty \rmd x
     \frac{ 8R^2(\rmi x + \delta\mu )} { 3 W^2 - 8R^2(\rmi x + \delta\mu)}
\end{equation}
where $R(z) = \sqrt{z^2 - W^2/4} - z$.

The result is particle-hole symmetric, namely $T_{\rm c}(x) = T_{\rm c}(1-x)$.
At $x\to 0$ and $x\to 1$, $T_{\rm c}$ diminishes because 
the ferromagnetism of this system is due to the kinetic energy of the
conduction electron.
In Fig.~\ref{FigTcJHinfty}, we also depict a curve 
$T_{\rm c}\propto x(1-x)$ proposed in ref.~\citen{Varma96}
for comparison.
This simple form roughly
reproduces the doping dependence of $T_{\rm c}$.
In the following section, we discuss more precisely about
the Curie temperature at finite $\JH/W$.

\subsection{Monte Carlo method for finite size clusters}

On a  finite-size clusters, it is possible to investigate
the double-exchange system  by numerical methods.
The result is unbiased and exact within the numerical errors.

The partition function of the present model with  localized
spins treated as classical rotators is defined by
\bequ
  Z = {\rm Tr_{\rm S}} {\rm Tr_{\rm F}}
    \exp \left( -\beta [ \Ham(\{ \vec m_i \}) - \mu \hat N]  \right),
\eequ
where ${\rm Tr_{\rm S}}$ and $ {\rm Tr_{\rm F}}$ represent
traces over spin and fermion degrees of freedom, respectively.
In the finite size system,
$Z$ is obtained by taking the trace over fermion degrees of freedom
first and spin degrees of freedom afterwards.
Fermion trace is directly calculated from the 
diagonalization of $2N \times 2N$ Hamiltonian matrix, where $N$
is the number of sites. Trace over spin degrees of freedom
is replaced by the Monte Carlo summation over spin configurations
$\{ \vec m_i \}$. 

For a fixed configuration of classical spins $\{ \vec m_i \} $,
the Hamiltonian is numerically diagonalized and we obtain
\beqarr
 && {\rm Tr_{\rm F}}
    \exp(-\beta [ \Ham(\{ \vec m_i \}) - \mu N])
  = \prod_{\nu=1}^{2N}
       \left[ 1 + \exp( -\beta( E_\nu(\{\vec m_i\}) - \mu ) ) \right].
   \label{TraceFermion}
\eeqarr
$E_\nu$ ($\nu=1,\ldots,2N$) are eigenvalues of the Hamiltonian
matrix for a given configuration $\{\vec m_i\}$.
We have the effective action for the classical spin system
\bequ
  S_{\rm eff}(\{\vec m_i\}) = - \sum_\nu
        \log \left( 1 + \rme^{-\beta( E_\nu - \mu ) } \right),
\eequ
which gives
\bequ
  Z = {\rm Tr_{\rm S}} \exp(-S_{\rm eff}).
\eequ
Monte Carlo update of spin configurations is performed 
 using the Boltzmann weight of the state 
$\{ \vec m_i \}$,
\begin{equation}
 P(\{ \vec m_i\}) \propto \exp(-S_{\rm eff}(\{\vec m_i\})).
\end{equation}
In a Monte Carlo unit step, orientations of
each spins are updated using the Metropolis algorithm.
Since the spins are classical, spin updates can be performed ergodically.

Thermodynamic quantities are stochastically calculated.
Quantities which are associated with localized spins are obtained
directly from the thermal average of spin configurations.
Electronic quantities are calculated from the
 eigenvalues and
eigenfunctions of $\Ham(\{ \vec m_i\})$.

One of the advantages for taking the classical spin limit is that
there exists no ``negative sign problem'', which is present
in quantum spin models. In the 
classical spin limit, the spin degrees of freedom
is completely decoupled from those of fermions, and the fermionic trace 
in eq.~(\ref{TraceFermion}) is obtained by solving the 
noninteracting lattice fermion system with random static potential.
Another advantage is that the real frequency dynamics of
electronic properties are directly obtained, since eigenvalues of the
Hamiltonian are calculated through the Monte Carlo procedures.
There is no difficulties of analytical continuations
form imaginary frequencies, as is present in some
quantum Monte Carlo methods.
A disadvantage of this method is that the auxiliary field is
static {\em i.e.} non-local in imaginary time, so it is not
possible to make a local spin flip using the imaginary time
Green's function as in the case for the Hubbard 
model.\cite{Hirsch81,Hirsch82}

\begin{figure}
\epsfxsize=15cm\epsfbox{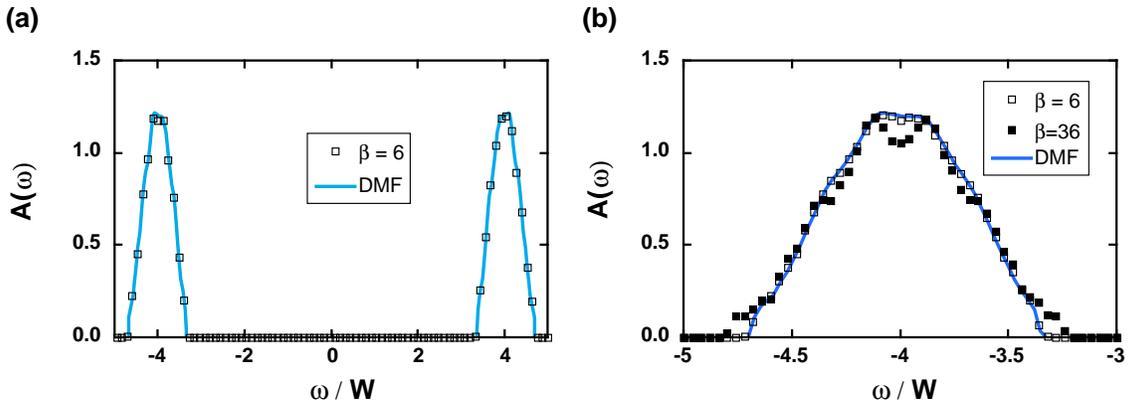}
\caption{Monte Carlo results for
DOS on $6\times 4\times 4$ cubic lattice at $\JH/W=4$,
at $\beta\equiv W/T=6$ and $36$. Error bars
are within the symbol size. Curves in the
figure show the DMF results. 
(left) Two peak structure at $\omega\sim \pm \JH$ is seen
for both DMF and Monte Carlo results.
(right) Lower subband at $\omega\sim -\JH$.
}
\label{FigClusterDOS}
\end{figure}

Let us now compare the dynamical mean-field theory
and the cluster Monte Carlo method.
Here we calculate the electron DOS
\bequ
   A(\omega) = - \frac1N \sum_{k\sigma}         
        \Im G_\sigma(k,\omega + \rmi \eta)/\pi
\eequ
on a finite size cluster system in Fig.~\ref{FigClusterDOS}.
We treat $N = 6\times 4\times 4$ 
cubic lattice at $\JH/W=4$ and $\mu=-\JH$, 
where the dynamical mean-field approach gives 
$T_{\rm c} = 0.028W$. In order to avoid
delta-function singularities in finite size systems,
 we use an adiabatic factor $\eta = 10^{-2}$ to smooth
the spectra.
In the Monte Carlo calculation,
we calculate at $\beta\equiv W/T=6$ and $36$ ,
 and compare with the dynamical mean-field
result in the paramagnetic phase $T > T_{\rm c}$.
The result shows that the DMF calculation is
very accurate.

\section{RESULTS}

In this section, we show the results for the 
classical spin limit $S=\infty$ of the DE system
using the DMF approach as well as the Monte Carlo calculation.
Hereafter, the electronic bandwidth 
is taken to be $W \equiv 1$ as a unit of energy.
For the carrier electron number, we
express by $x = 1 - \brakets{n}$.
We define $M = \braket{m_z}$, where $0\le M \le 1$.
Also, we describe the total moment, or sum of moments of
localized spin and electron spin, as
$M_{\rm tot} = \braket { \frac32 m_z + \frac12 \sigma_z}$.
In a normalized form we describe $M^* = M_{\rm tot} / M_{\rm sat}$,
where $M_{\rm sat}$ is the saturation value of $M_{\rm tot}$ at
the ground state.
We make distinctions between $M$ and $M^*$, since
analytical calculation is better understood by $M$ while
the comparison with experiments should be done by $M^*$.
Nevertheless, in the strong coupling region $\JH\gg W$
we have $M\simeq M^*$ so effectively there exists no major
differences.


\subsection{Electronic structures}
\label{DOSstructure}
\subsubsection{Spectral function and the density of states.}
In Fig.~\ref{FigAw} we show the spectral function 
for the up-spin electron $A_\uparrow(k,\omega)$
 on a cubic lattice, where $k/\pi = (\zeta,\zeta,\zeta)$.
From the particle-hole symmetry and the spin symmetry,
down-spin part $A_\downarrow(k,\omega)$ is reproduced by
the relation 
\begin{equation}
  A_\downarrow(k,\omega) = A_\uparrow(Q-k,-\omega),
\end{equation}
where $Q=(\pi,\pi,\pi)$.

\begin{figure}[htb]
\epsfxsize=15cm\epsfbox{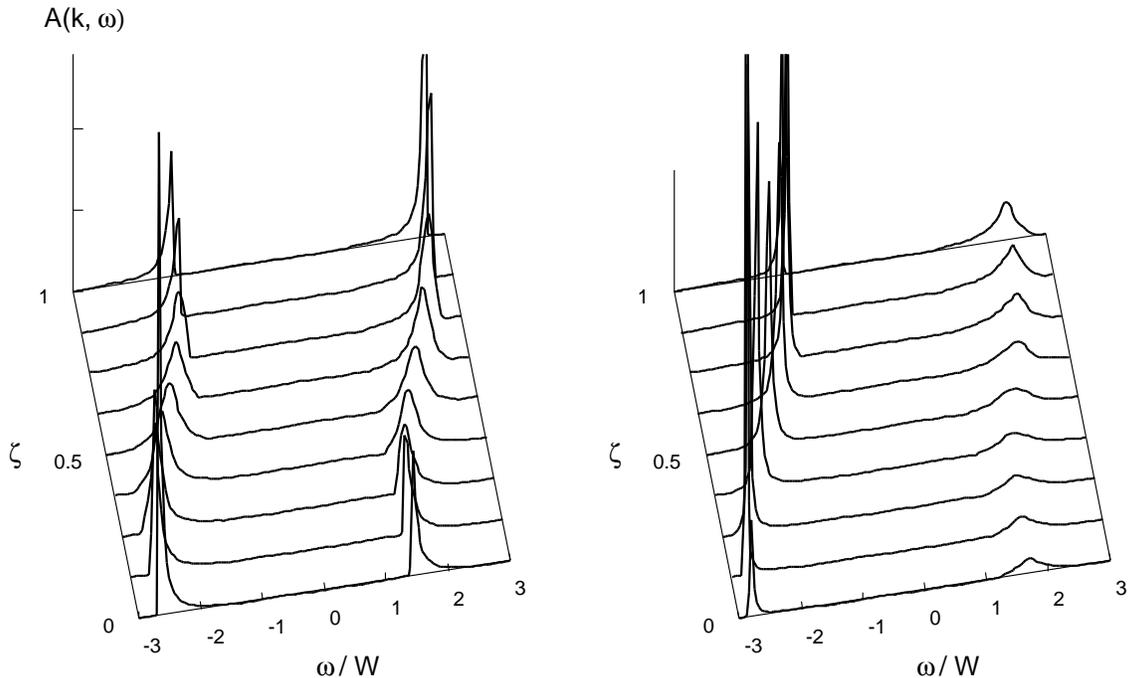}
\caption{Spectral function $A_\uparrow(k,\omega)$ 
on a cubic lattice at $\JH/W=2$ and
x=0.3, (left) in the paramagnetic phase $T=1.05T_{\rm c}$, and
(right) in the ferromagnetic phase $T=0.5T_{\rm c}$.
Here, $k/\pi = (\zeta,\zeta,\zeta)$.}

\label{FigAw}
\end{figure}

In Fig.~\ref{FigAw} we show the quasiparticle excitation structure
and its temperature dependence.
There exists
two-peak structure at around $\omega \sim \pm \JH$.
Above $T_{\rm c}$, peaks at upper and lower bands are
symmetric and equally weighted.
Below $T_{\rm c}$, the structure remains split but becomes asymmetric.
For the up-spin electron, the integrated weight is transferred from 
upper band to lower band. 
We also see the change of the quasiparticle linewidth $\Gamma$.
The lower band peak becomes sharper, which means the
reduction of $\Gamma$ or the enlonged quasiparticle lifetime.
 On the other hand,
$\Gamma$ for the upper band peak increases.

To investigate the change of the spectral weight in further detail,
we calculate the DOS $A_\sigma(\omega)$ by $k$-integrating the spectral weight.
In Fig.~\ref{FigDOS} we show the DOS as a function of temperature
in the paramagnetic and  ferromagnetic phases.
Two subband structure at  $\omega\sim \pm \JH$ reflects
the quasiparticle structure as described above.
At the ground state, lower subband is composed of up-spin
only, and down-spin band exists only at the high-energy region.
The bandwidth of the DOS becomes narrower as temperature becomes higher.
The band center is fixed at $\omega\sim \pm \JH$ which is
the energy level of the atomic limit $t=0$.

\begin{figure}[htb]
\epsfxsize=12cm\epsfbox{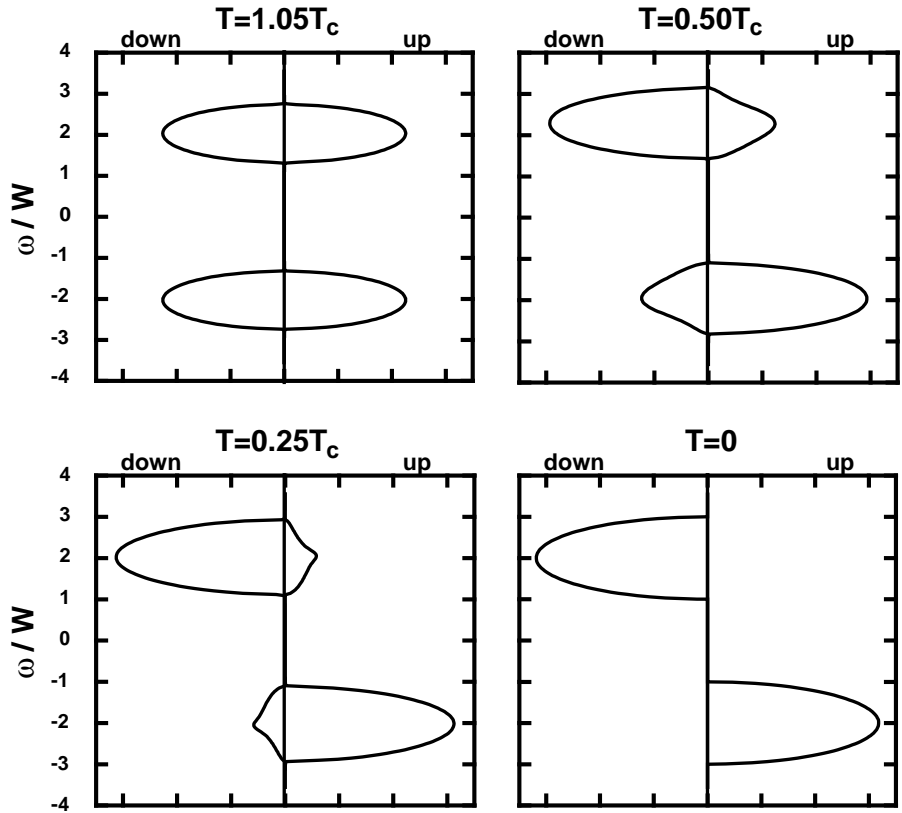}
\caption{Temperature dependence of DOS for $\JH/W=2$ and $x=0.3$,
where $T_{\rm c} = 0.019W$. Peak structures are observed at 
$\omega\sim \pm\JH$.}
\label{FigDOS}
\end{figure}

These results are in agreement with the exact result in a limiting case
(Lorentzian DOS with $\JH\to\infty$) discussed in 
\S\ref{DMFanalytical}. Let us focus on the lower subband.
The DOS is nearly proportional to the spin polarization of the
localized spin, namely
\begin{eqnarray}
  A_\uparrow(\omega) & \sim  z_\uparrow^{(l)}(M) &= (1+M)/2,
   \nonumber\\
  A_\downarrow(\omega) & \sim  z_\downarrow^{(l)}(M) &= (1-M)/2,
\end{eqnarray}
where $z$ is the quasiparticle weight discussed in \S\ref{DMFanalytical}.
We see that $z$ is determined by the magnetization through $ (1\pm M)/2$,
which is the probability that the local spin is parallel 
to the itinerant electron with up (down) spin. This is easily understood
from the nature of the double-exchange interaction which
projects out the antiparallel component of the spins.
This spin-dependent projection
may be viewed as the evolution of the majority-minority band structure.
 The up (down) spin band becomes a majority (minority) band
below $T_{\rm c}$ or under magnetic field.
Shift of the spectral weight from the
minority band to the majority band occurs 
as temperature is lowered. At the ground state ($M=1$), minority band
completely loses its weight.

Change of the electron bandwidth is understood
qualitatively through Anderson-Hasegawa's picture.\cite{Anderson55}
Electron hopping amplitude is proportional to 
$ \cos(\theta/2)$ where $\theta$ is the relative angle 
of the localized spins.
At high temperature, 
$\theta$ deviates from zero due to spin fluctuation,
and the amplitude of  electron hopping matrix element 
and hence the bandwidth decreases.
This is also shown by
the virtual crystal approximation.\cite{Kubo72a}

In order to account for
 the width of $A(k,\omega)$, or the quasiparticle lifetime,
we have to go beyond a mean-field picture like in Anderson-Hasegawa
approach.
The origin of the linewidth $\Gamma$ is the thermal fluctuation of the spins.
In the strong coupling region $\JH \gg W$, spin scattering
phenomena in the paramagnetic phase is so large that quasi-particles 
lose their coherence due to the
inelastic scattering by thermally fluctuating spins.
In the ferromagnetic phase, the spin fluctuation decreases
as temperature is decreased.
For the majority band, this decreases $\Gamma$,
and in the limit $T\to0$ the majority band becomes a free electron band.
However, for the minority band, the spin projection causes further
loss of the coherence, which leads to the increase of $\Gamma$
 as well as the decrease of the quasiparticle weight $z$.
Asymptotically in the limit $T\to 0$, $\Gamma$
approaches to a constant $\sim W$, and at the same time $z\to0$.
Then, at $T=0$, the minority band with finite $\Gamma$ is projected out.

\subsubsection{Half metal.}
Metal with a DOS structure shown in Fig.~\ref{FigDOS} where
only one of the spin species have the Fermi surface is called
a half metal.\cite{deGroot83,Irkhin94}
Namely, because of the `Zeeman splitting' due to Strong Hund's coupling
 $\JH\gg W$,
ferromagnetic ground state of the DE model 
shows a perfect spin polarization and thus is a half-metal.
Experimentally, spin-resolved photoemission investigation\cite{Park98a,Park98b}
shows that the conduction band of the doped manganites is a half-metal.
Artificial trilayer junction of 
manganites\cite{Sun96} also shows a large tunneling magnetoresistance
phenomena, and the spin polarization is estimated to be more than 80\%.
Such a DOS structure creates a phenomena called
tunneling magnetoresistance (TMR), which will be
 discussed in \S\ref{Experimental}.


\subsubsection{Shift of the chemical potential.}
A direct consequence of the change of the bandwidth controlled
by the magnetization will be observed in the shift of the chemical
potential.
In this case, the change of the DOS structure is in a way such that
the band center is pinned by the Hund's coupling energy $\pm \JH$
and the band edge shifts away from the center as magnetization
is increased. Then, for a hole doped case the position of the
chemical potential increases by increasing the magnetization.

\begin{figure}[htb]
\epsfxsize=8cm\epsfbox{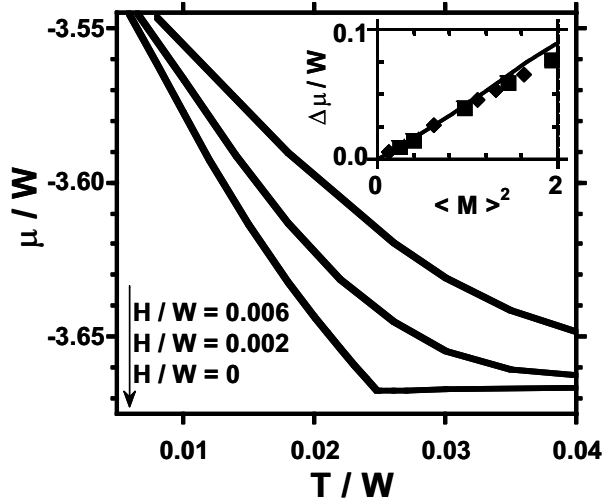}
\caption{Temperature dependence of $\mu$ at 
$\JH/W=4$ and $x=0.20$ under
various magnetic field.
Inset: $\Delta\mu/W$ as a function of $M^2$.
Lines show the result at $T<T_{\rm c}$ for $H=0$.
Squares and diamonds are data at $T=1.01 T_{\rm c}$ and $1.2 T_{\rm c}$
by applying $H$,
respectively.
}
\label{FigT-M-mu}
\end{figure}

In Fig.~\ref{FigT-M-mu} we show the temperature dependence
of the chemical potential at $x=0.2$
 under various magnetic field.\cite{Furukawa97}
At $H=0$, chemical potential $\mu$
 is nearly temperature independent above $T_{\rm c}$.
Below $T_{\rm c}$, $\mu$  shifts as a function of
temperature.
We also calculate $\mu$ and 
$M$, (i) at $H=0$ by changing temperature
 in the region $T \le T_{\rm c}$,
and (ii) at fixed temperature above $T_{\rm c}$ by changing $H$.
In the inset of Fig.~\ref{FigT-M-mu} we plot
$\mu$ as a function of magnetic moment $M^2$ for both cases.
As a result, we see the scaling relation
\begin{equation}
  \Delta\mu/W \propto  M^2,
     \label{ScalingRelation}
\end{equation}
where $\Delta\mu \equiv \mu(T,H) - \mu(T=T_{\rm c},H=0)$.
We see that $\Delta\mu$ can be as large as $0.1W$.

Thus, for a fixed band filling,  the total change of the
DOS width in the entire energy range 
causes the shift of $\mu $.
The change in such a large energy scale controlled by
magnetization produce the
characteristic feature of the shift of $\mu$ in DE systems;
namely, that the shift of $\mu$ is as large as a few tenth of $W$
and that the scaling relation (\ref{ScalingRelation}) is
satisfied up to such a large energy scale.

Such a large shift of $\mu$ might possibly be applied to
electronic devices which controls the MOS gate voltage by
the magnetic field.

\subsection{Magnetic structure and transport properties}

\subsubsection{Magnetic transition temperature.}
In the limit $\JH\to\infty$,
Curie temperature  $T_{\rm c}$ of the DE model is
determined by the electron kinetic energy.
Indeed, DMF calculation shows 
that $T_{\rm c}$ is scaled by
electron hopping, {\em i.e.} $T_{\rm c} \propto W$ for
$\JH=\infty$ limit (see \S\ref{DMFanalytical}).
Here we show the case for finite $\JH/W$.

In Fig.~\ref{FigTCurie} we show the Curie temperature $T_{\rm c}$
as a function of doping $x$ for various values of $\JH/W$.
At finite $\JH/W$, $T_{\rm c}$ is reduced
from $\JH=\infty$ values. We also see that
$T_{\rm c}$ systematically increase
as $x$ is increased and have maximum at around $x\sim 0.5$,
which is due to the increase of the kinetic energy.

\begin{figure}[htb]
\epsfxsize=8cm\epsfbox{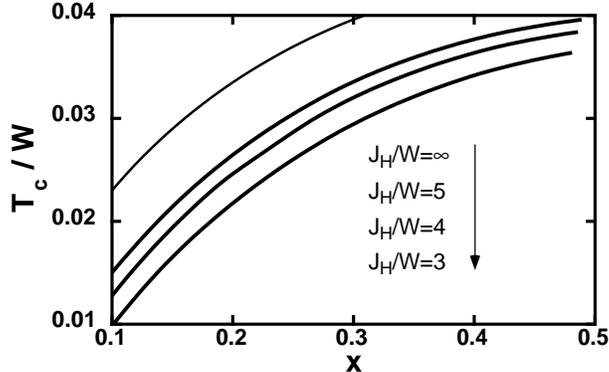}
\caption{Curie temperature $T_{\rmc}$
 as a function of $\JH/W$ and $x$.}
\label{FigTCurie}
\end{figure}

At half-filling $x=0$, there exists antiferromagnetic order.
In Fig.~\ref{FigNeel}
we show the phase diagram at $x=0$ for the $D=\infty$ Bethe lattice.
In the weak coupling region $\JH \ll W$, the N\`eel temperature $T_{\rm N}$
is equivalent to the results from the SDW mean-field type equation
with $\JH/W$ dependence in an essential singular function.
At $\JH\gg W$, $T_{\rm N}$ is determined from the Heisenberg model
with the exchange coupling $J_{\rm AF} \sim t^2/\JH \propto W^2/\JH $.

\def\Neel{N\`eel}

\begin{figure}[htb]
\epsfxsize=8cm\epsfbox{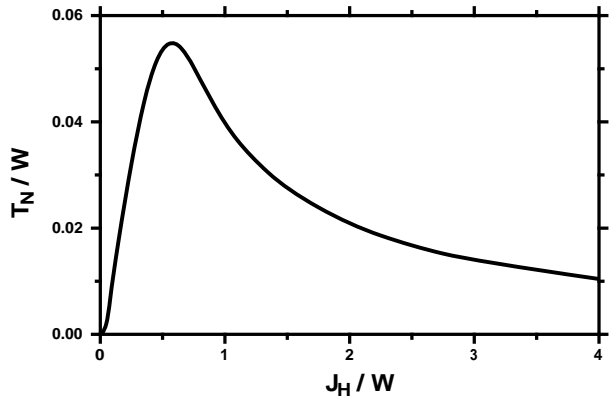}
\caption{N\`eel temperature at $x=0$.}
\label{FigNeel}
\end{figure}

Thus we see the antiferromagnetic order at $x=0$ and 
ferromagnetic ground state at sufficiently doped case.
In the underdoped region $x\ll 1$, de Gennes discussed the
existence of the canted state \cite{deGennes60}.
However, in \S\ref{PhaseSeparation} we show that
the canted state is unstable against phase separation,
{\em i.e.} mixed phase of $x=0$ antiferromagnetic region and
$x >0$ ferromagnetic region.\cite{Yunoki98}


\subsubsection{Resistivity as a function of magnetization.}
Resistivity $\rho(T)$ as well as total magnetization $M_{\rm tot}$
 as a function of temperature is given in Fig.~\ref{FigTempDep}.
Here, 
$\rho_0$ is a constant of resistivity which corresponds
to the Mott's limit value (inverse of the Mott's minimum
conductivity) in three dimension.
$M_{\rm sat}$ is the saturated magnetization at $T=0$.

\begin{figure}[htb]
\epsfxsize=8cm\epsfbox{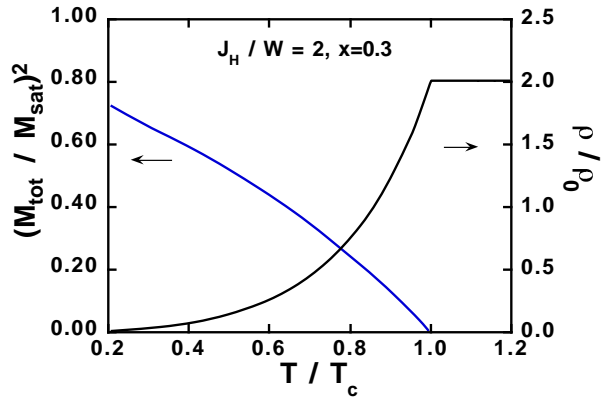}
\caption{Resistivity $\rho$ and magnetization $M_{\rm tot}$
 as a function of temperature.}
\label{FigTempDep}
\end{figure}

Above $T_{\rm c}$, the value of the resistivity is
in the order of the Mott limit $\rho(T) \sim \rho_0$,
with small $T$-dependence.
Below $T_{\rm c}$, resistivity drops quickly as
magnetization increases.
From the DMF 
calculation,\cite{Furukawa94,Furukawa95d}
it has been made clear that the resistivity behaves as
\begin{equation}
  \rho(M)/ \rho(M=0) = 1- C M^2,
   \label{RhoScaleTheory}
\end{equation}
where $C$ is a temperature/field independent 
constant.
Namely, temperature and magnetic field dependences 
come from the magnetization $M=M(H,T)$. As a function of magnetization $M$,
all the $T$ and $H$ dependent values $\rho(T,H)$ converge on a
 universal curve in (\ref{RhoScaleTheory}).
In other words, the origin of the resistivity 
is due to spin fluctuation, or more precisely spin disorder scattering,
discussed by Kasuya\cite{Kasuya56} and later by
Fisher and Langer.\cite{Fisher68}

In the Born approximation (weak coupling limit), we see $C=1$.\cite{Kasuya56}
A phenomenological treatment by Kubo and Ohata which
estimates the resistivity from the spin fluctuation,
\begin{equation}
  \rho \propto (\delta S)^2 \propto 1- M^2,
\end{equation}
 also 
gives $C=1$. However, in the DMF at $\JH\gg W$, we have $C>1$
which indicates the strong coupling behavior.
In the next section,
the relation with experimental MR is discussed in more detail.

Above $T_{\rm c}$,  small $T$-dependence in $\rho(T)$
is observed within the DMF treatment.
This result might be an artifact of the approximation
since the local spin fluctuation is saturated above $T_{\rm c}$.


\subsection{Charge and spin dynamics}
\label{DEdynamics}

\subsubsection{Optical conductivity.}
Temperature dependence of the optical conductivity is shown in
Fig.~\ref{FigOpt}.
In the paramagnetic phase, the spectrum splits into
two peaks due to the 2-subband structure of the DOS.
Namely, intraband particle-hole channel creates a
Drude-like peak at $\omega\sim 0$, while interband channel
creates a peak at around $\omega\sim 2\JH$.
In the inset of Fig.~\ref{FigOpt}, we show the 
weight of the interband process as a function of $1-M^*{}^2$.
The integrated weight of the interband optical process at $\omega\sim 2\JH$
defined by
\begin{equation}
  S = \int_{\omega_{\rm c}}^{\infty}{\rm d}\omega\  \sigma(\omega),
\end{equation}
where cutoff frequency is taken as $\omega_{\rm c}=\JH$.
We see a scaling relation
\begin{equation}
  S \propto 1-M^*{}^2.
   \label{OCscaleTheory}
\end{equation}

\begin{figure}[htb]
\epsfxsize=8cm\epsfbox{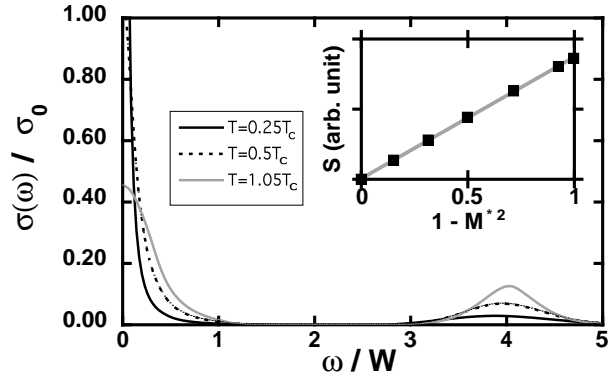}
\caption{Temperature dependence of the optical conductivity
for $\JH/W=2$ and $x=0.3$.
Inset: Integrated weight $S$ at high energy part $\omega\sim 2\JH$.}
\label{FigOpt}

\end{figure}

This is explained by the temperature dependence of DOS.
Below Curie temperature, the DOS changes as in Fig.~\ref{FigDOS}.
Interband optical process at $\omega\sim 2\JH$
is constructed from a process of 
making a pair of lowerband hole and upperband electron.
Then, the transfer of spectral weight by magnetization 
creates the change in the optical spectra as follows.
For the up spin electrons, the spectral weight of lower and upper subbands
are proportional to  $(1+M^*)/2$ and $(1-M^*)/2$, respectively.
Since this optical process conserves quasiparticle spin,
the total weight of the optical spectrum is
proportional to the product of  the initial-state weight
at lower subband and the final-state
weight at the upper subband, and hence  scaled as $1-M^*{}^2$.
Contribution from the down-spin band is also the same, and we have
$S \propto 1-M^*{}^2$.

\subsubsection{Stoner excitation.}
Stoner susceptibility is calculated by
\begin{equation}
  \chi(q,z) = \frac1{\beta N}\sum G(k+q,\rmi\omega_n + z ) G(k,\rmi\omega_n).
\end{equation}
Here, correlation effects are taken into account through
the self-energy correction in $G$.
In Fig.~\ref{FigImChiAll} we show 
$q$-dependence of ${\rm Im}\,\chi(q,\omega)$
 for various temperatures,
at $\JH/W=2$ and  $x=0.3$.
We see two-peak structure at $\omega\sim 0$ and $\omega\sim 2\JH$ which
is explained from the $\JH$-split DOS structure.
The Stoner absorption is produced from a particle-hole 
pair excitations with spin flip, which produces a peak at low energy
from intra-band processes and another peak at $\omega\sim2\JH$
from interband processes.

We see a weak $q$ dependence in the low frequecy part,
especially at $T\ll T_{\rm c}$. This part of the Stoner process
is dominated by the combination of the majority-minority quasiparticles.
Since the quasiparticles in the minority band is incoherent,
$\Im\chi(q,\omega)$ is weakly $q$ dependent.
On the other hand, high energy part of $\Im\chi(q,\omega)$
at $T\ll T_{\rm c}$ have larger $q$ dependence. 
This part is dominated by the majority-majority quasiparticle channel.
Thus $q$ dependence of $\Im\chi(q,\omega)$ reflects the band structure
of the quasiparticles.

\begin{figure}[htb]
\epsfxsize=8cm\epsfbox{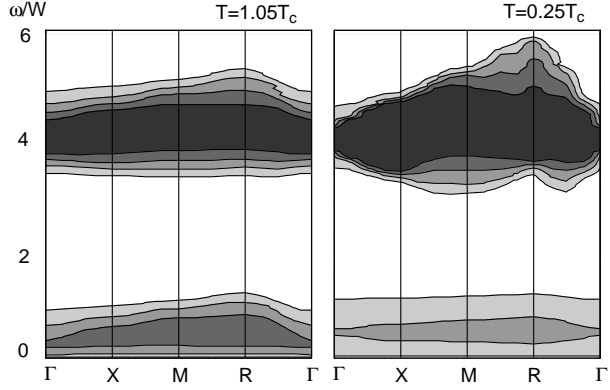}
\caption{Contour plot of the Stoner absorption  $\chi(q,\omega)$ 
on a cubic lattice at $\JH/W=2$ and $x=0.3$, in 
the paramagnetic phase $T=1.05T_{\rm c}$ (left)
 and in the ferromagnetic phase $T=0.25T_{\rm c}$ (right).
At $T=0$, the low energy part at $\omega\sim0$ disappears.}
\label{FigImChiAll}
\end{figure}

Let us see the $\omega$ dependence at the low frequency region.
In Fig.~\ref{FigImChiZB} we show $\chi_{\rm ZB}(\omega) = \chi(Q,\omega)$
where $Q=(\pi,\pi,\pi)$.
We see that at small $\omega$ we have $\omega$-linear
relation, {\em i.e.} 
$  \Im\chi \propto \omega$ at $\omega \ll W$.
Coefficients for $\omega$-linear part
  decrease by decreasing the temperature, and
 we find\cite{Furukawa98}
\begin{equation}
  \Im\chi(Q,\omega) \propto (1-{M^*}^2) \,\omega
   \label{ImChiScale}
\end{equation}
for small values of $\omega$.
The relation (\ref{ImChiScale})
is observed at all values of $q$ with weak $q$ dependence.

\begin{figure}[htb]
\epsfxsize=8cm\epsfbox{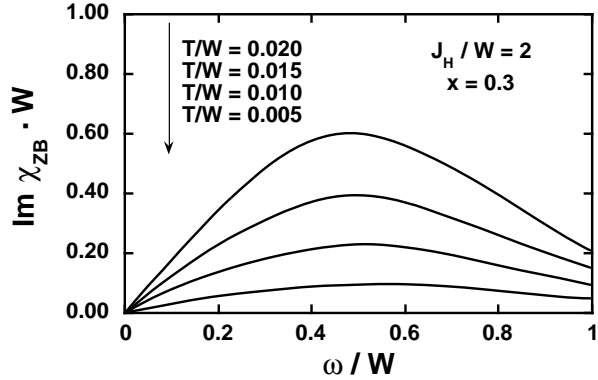}
\caption{Stoner absorption $\chi(Q,\omega)$ at the zone boundary
$Q=(\pi,\pi,\pi)$. At $\JH/W=2$ and $x=0.3$, transition
is at $T_{\rm c}/W = 0.019$.}
\label{FigImChiZB}
\end{figure}

Low energy Stoner absorption is
constructed from a minority particle and majority
hole channel.
Since the majority and the minority bands have the spectral weight
proportional to $(1+M^*)/2$ and $(1-M^*)/2$, respectively,
the low energy part of the Stoner absorption is proportional to
their product, $1-M^*{}^2$. 
The $\omega$-linear behavior comes from the Fermi distribution
function.
The incoherence of the minority band gives the weak $q$ dependence.
 Thus we have the scaling relation
$\Im\chi \propto (1-{M^*}^2) \omega$.
The weak $q$ dependence is in large contrast with the 
conventional weak ferromagnet where minority band is also coherent,
which gives strong $q$ dependence through its band structure.


\subsection[]{Phase separation\footnote{Works shown here 
concerning the issue of phase separation has
been made in collaboration with
S. Yunoki, A. Moreo and E. Dagotto.}}
\label{PhaseSeparation}

Magnetic phase diagram of the weakly doped DE model has been studied
by de Gennes.\cite{deGennes60} Assuming the homogeneity of the
doped carriers, he concluded that the spin canted phase is the most
energetically favorable state.
However, we have recently shown that there exists an instability
toward phase separation,\cite{Yunoki98} and  the assumption of uniformly doped
charges by de Gennes is not valid.

One of the ways to discuss the phase separation is to
make a grand canonical calculation of the particle number $x$ as a function
of the chemical potential $\mu$. 
If there exists a jump of $x(\mu)$ at the critical value
$\mu=\mu_{\rm c}$ in the thermodynamic limit, it implies
that two phases with different doping $x$ coexist at $\mu=\mu_{\rm c}$.
We have shown\cite{Yunoki98} the jump of $x(\mu)$ in the DE model
for sufficiently large $\JH/W$. The jump occurs from 
$x=0$  to a state with finite $x$.
The calculation has been performed within the DMF approach ($D=\infty$),
and cross checked by the Monte Carlo method for 
the $D=1$ and $2$ clusters in the extrapolated limit of $T\to0$.

Phase boundary is simply determined from the jump of $x(\mu)$.
In Fig.~\ref{FigPhaseDiagramDoped} we show the 
$x$-$T$ phase diagram. At the low temperature region,
we see mixed phases of $x=0$ AF state and doped ($x>0$) state
with either paramagnetic or ferromagnetic state,
depending on temperature.
\begin{figure}[htb]
\epsfxsize=8cm\epsfbox{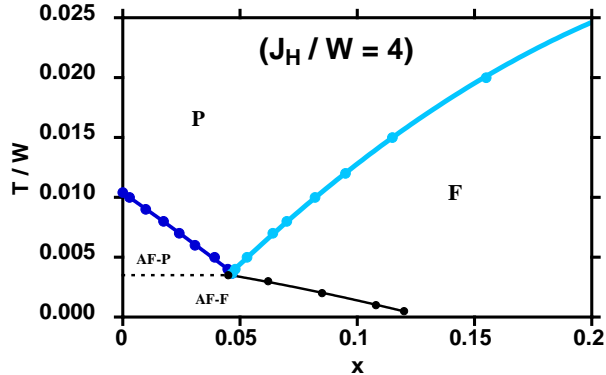}
\caption{Phase diagram at $\JH/W=4$. P (F) represents paramagnetic
 (ferromagnetic) region. At $x=0$ we have antiferromagnetic (AF) phase.
Regions labeled by AF-P (AF-F) are phase separated region with
mixed phases of AF and P (AF and F) phases.
}
\label{FigPhaseDiagramDoped}
\end{figure}

The mechanism of the phase separation may be understood as follows.
In Fig.~\ref{FigDOSpfaf} we show the change of the density of states
for different doping concentration.
We see that the bandwidth substantially differs for different
magnetic states, while the band center remains at $\omega\sim\pm \JH$.
As discussed in \S\ref{DOSstructure}, this typical DOS structure
is a consequence of the DE half-metallic system.
Let us consider the zero temperature limit. At the
$x=0$ AF state, we have $\mu=0$. In order to hole dope this
AF state, we need to decrease $\mu$ to a AF gap-value,
$\mu=-\Delta_{\rm AF}$. 
However, as we see in Fig.~\ref{FigDOSpfaf}, it is also
possible to make a  F state at $\mu=-\Delta_{\rm AF}$ with
rich hole density $x$. AF state gains the exchange energy $\sim t^2/\JH$,
while the F state gains the kinetic energy $\sim t x$.
Thus, in the limit $\JH\gg t$, the doped F state becomes energetically
favorable. In such a case, there exists the chemical potential $\mu_{\rm c}$
such that $-\Delta_{\rm AF} < \mu_{\rm c} < 0$, where
the energies of AF state with $x=0$ and F state with $x>0$ are equal.
Level crossing from AF to F states occurs at this
critical point $\mu=\mu_{\rm c}$.
Doped F state is realized before doping the AF state.
Thus we have a jump in the hole density from $x=0$ to finite $x$.

\begin{figure}[htb]
\epsfxsize=8cm\epsfbox{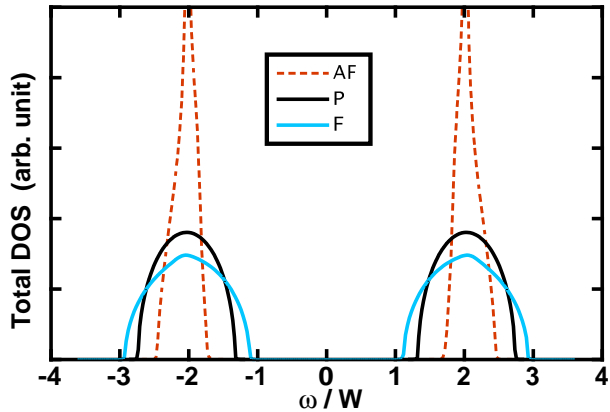}
\caption{Density of states for  ferromagnetic (F), paramagnetic (P)
and antiferromagnetic (AF) states at $x=0.2$, $0.1$ and $0$, respectively,
for $\JH/W=2$ and $T/W=0.005$.}
\label{FigDOSpfaf}
\end{figure}

The discussion based on the DOS structure is quite generic.
In strongly correlated systems with bistable phases,
macroscopic change of the order parameters creates
a large change in the electronic DOS.
This gives a macroscopic jump in $\braket{n}$ from one phase to another
when $\mu$ is fixed. Then, phase separation is associated with
the density-driven phase transition.

Let us discuss the issue from a different viewpoint.
In Fig.~\ref{FigClusterSq} we show the equal time spin correlation
\bequ
  S(q) = \frac1N \sum_{ij} \brakets{ \vec S_i \cdot \vec S_j}
		\rme ^{\rmi q (i-j)}
\eequ
on a one dimensional system at $L=40$, $\JH/W=4$ and $\beta=150 W^{-1}$.
We clearly see the crossover from the antiferromagnetic state at $n=1$ 
with a peak of $S(q)$ at $q=\pi$,
and the ferromagnetic state at $n\sim 0.7$ with the peak at $q=0$.
Around $n\sim 0.9$, where large change  of $n(\mu)$ is observed,
we see two peak structure at $q=\pi$ and $q=0$. 
This result is consistent with mixed phase of 
ferromagnetic and antiferromagnetic states.
On the other hand, magnetic states with incommensurate momentum,
typically at $q=2k_{\rm F}$, have been observed
for the weak coupling region $\JH\simle W$. 
Thus the tendency toward phase separation 
is prominent at the strong coupling region.

\begin{figure}[htb]
\epsfxsize=8cm\epsfbox{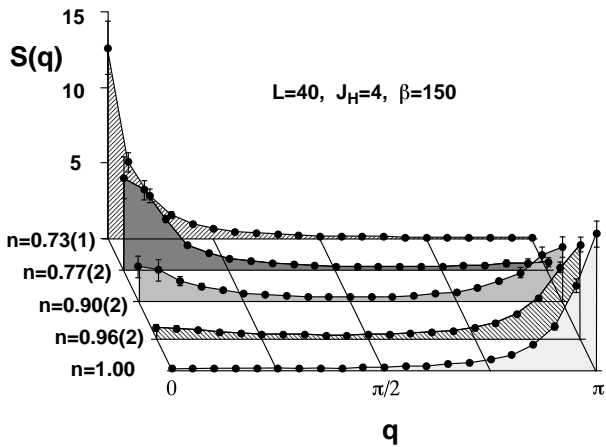}
\caption{Equal time spin correlation for various electron concentration.
Shaded area are guides to eyes. Error bars of the concentrations 
in the last digits are given in parenthesis.}
\label{FigClusterSq}
\end{figure}

Recently, the issue of phase separation is also investigated by various 
methods.\cite{Riera97,Kagan98x,Arovas98x}
The model in the weak coupling region $\JH \ll W$ has also
been studied.\cite{Nagaev97} However, the mechanism of the phase 
separation might be different from the strong coupling region 
where the half-metallic DOS plays an important role.
In manganites, several experiments claim the existence of
phase separation.\cite{Allodi97,Hennion98}
We will later discuss this issue in comparison with experiments in detail.

\section{1/S CORRECTIONS}
\label{SW}
\subsubsection{Spin wave expansion.}

We introduce a linear spin wave theory.
The spin wave operators are introduced from
\begin{equation}
   S_i{}^+ \simeq \sqrt{2S} a_i, 
\qquad
   S_i{}^- \simeq \sqrt{2S} a_i\dags, 
\qquad
   S_i{}^z = S - a_i\dags a_i.
		\label{defSWope}
\end{equation}
Hereafter we  restrict ourselves to 
the lowest order terms of the $1/S$ expansion at $T=0$, where
 the localized spins are perfectly polarized
so that $\brakets{a_i\dags a_i} = 0$.
We consider the half-metallic ground state, {\em i.e.} $f_{k\downarrow}=0$
where $f_{k\sigma}$ is the Fermi distribution function.

The spin wave self-energy in the lowest order of $1/S$ expansion
is obtained diagramatically as\cite{Furukawa96}
\begin{eqnarray}
    \Pi(q,\omega) &=&
	 -  \frac{2\JH^2}{S} \frac1{N\beta}
		\sum_{k,n}	
	G_\uparrow(k,\rmi \omega_n) G_\downarrow(k+q,\rmi\omega_n + \rmi\nu)
	\nonumber	\\
   &&
	-\frac{\JH}{S} \frac1{N\beta}
     \sum_{k,n}
		\left(
    G_\uparrow(k,\rmi \omega_n) - G_\downarrow(k,\rmi\omega_n)
		\right) \rme^{\rmi \omega_n 0_+} 
	\nonumber	\\
 &=& \frac{1}{SN}
    \sum_k f_{k\uparrow}
	\left(\JH -  \frac{2\JH^2}
		{ 2\JH - (
	  \omega + \varepsilon_k - \varepsilon_{k+q} ) } 
	\right).
	\label{defPiz}
\end{eqnarray}
Here, $G_\sigma(k,\rmi\omega_n)= 
(\rmi\omega_n - \varepsilon_k + \sigma \JH)^{-1}$
 is the fermion Green's function, and $\beta=1/T$.

The spin wave dispersion relation $\omega_q$ is obtained self-consistently
from $  \omega_q = \Pi(q,\omega_q)$.
We have
\bequ
  \omega_q = \frac{1}{SN}
    \sum_k f_{k\uparrow}
	\left(\JH -  \frac{2\JH^2}
		{ 2\JH - ( \varepsilon_k - \varepsilon_{k+q} ) } 
	\right) + O(1/S^2),
    \label{defDispRel}
\eequ
where $f_{k\uparrow}$ is the fermi distribution function
of the majority band.

Let us consider  the strong Hund coupling limit $\JH \gg t$.
If we assume a simple cubic lattice with nearest-neighbor
electron hopping,
\bequ
  \varepsilon_k = -2t \left(
	\cos k_x + \cos k_y + \cos k_z
	\right),
		\label{def DispRelFermi cubic}
\eequ
we have
\bequ
  \omega_q \simeq E_{\rm sw}
	\frac{3 - \cos q_x - \cos q_y - \cos q_z }{6}
	\label{SW, allBZ, nn}
\eequ
where $E_{\rm sw} \equiv \omega_{q=Q} - \omega_{q=0}$ 
is the spin wave bandwidth, given by
\bequ
  E_{\rm sw} = \frac{6t}{SN} \sum_k f_{k\uparrow} \cos k_x .
\eequ
We see that in the strong coupling region the spin wave bandwidth
is determined only by the electron transfer energy.

In the isotropic case, the spin stiffness is defined via
$\omega_q = Dq^2$ in the long wavelength limit $q\to0$.
From eq.~(\ref{defDispRel}) we have
\bequ
   D = \frac1{2S}
		\frac1N\sum_k f_{k\uparrow}
    \left[
         \frac12 \frac{\partial^2 \varepsilon_k} {\partial k^2}
         -   \frac{  1}{2\JH} 
	    \left(
	     \frac{\partial \varepsilon_k} {\partial k}
	    \right)^2
    \right].
		\label{defDmunu}
\eequ

We emphasize that the expansion given here is not
with respect to $\JH/(tS)$ but to $1/S$, due to energy denominator
$2\JH$ between up- and down- spin electrons.
This is understood from the fact
 that $\JH$ plays an role of a projection
and does not enter the energy scale by itself in the limit $\JH\to\infty$.
Therefore, the calculation is valid 
even in the large Hund's coupling limit 
$\JH \gg t$ as long as we restrict ourselves to $T=0$.
Indeed, the result for $\JH\to\infty $ obtained by the present
approach is equal to those in the projection limit $\JH=\infty$
shown by Kubo and Ohata.\cite{Kubo72}

\subsubsection{Discussion.}
In the DE model with $\JH \gg W$ we observe a short-range spin interaction,
in contradiction with the case of $\JH \ll W$ ($s$-$d$ model)
where the well-known RKKY interaction is long ranged
with a power-law decay.
The difference comes from the electronic structure.

In the case $\JH\ll W$ where electronic DOS is the ordinary one
(not half-metal), the gapless quasi-particle excitation 
of the electronic part creates a particle-hole spin excitation channel
with $2k_{\rm F}$ singularities. Interactions among localized spins
are mediated by this gapless channel and thus long ranged.

On the other hand, for the DE model with $\JH\gg W$ the
half-metallic DOS creates a gap due to $\JH$ splitting,
and unlike the former case the particle-hole spin channel is
massive and short ranged.
The qualitative difference comes from the half-metallic structure.
In the real-space picture, 
we consider a perfectly spin polarized state at $T=0$ and
twist a spin at site $i_0$. 
Spin polarization of itinerant electrons are along the total
polarization axis except for the $i_0$-th site where it
orients toward the local spin direction $\vec S_{i_0}$.
 In the strong coupling limit
$\JH \gg W$, the electron at site $i_0$ is localized because
it has different spin orientation from the spins in neighboring sites.
Since the effective interaction between localized spins
are mediated by the motion of electrons,
the effective spin-spin interaction is short ranged.
As $\JH/W$ increases, electrons become more localized so
the range of effective interaction becomes shorter.
In the extreme limit $\JH\to\infty$, the interaction
is nonzero only for the nearest neighbors which gives
a cosine-band dispersion.

Let us discuss the higher orders of  $1/S$ expansion terms.
In an ordinary (insulating) Heisenberg ferromagnet,
the first $1/S$ term is the relevant term with respect to
the one-magnon dispersion $\omega_q$. Higher orders of $1/S$ expansion
only give magnon-magnon interaction terms, and thus irrelevant within
the one magnon Hilbert space.
However, for the DE model, higher order terms also give
one-magnon kinetic terms. We need to take into account
the asymptotic $1/S$ expansions even for the one-magnon dispersion
relations. Such higher order terms may be considered as
vertex corrections to the self-energy term $\Pi(q,\omega)$.

Numerically, Kaplan {\em et al.}\cite{Kaplan97} studied the
$S=1/2$ case at $\JH\to\infty$. They observed  cosine-band type
behaviors in the well-doped cases, and the deviation from them
in the limit $n\to 0$ and $n\to1$. The result might be understood
from the Migdal's discussion. Let us consider
 the electron kinetic energy $E_{\rm kin}$
and the energy scale of magnons $\braket{\omega_q}$ 
of the DE model.
 The deviation from a cosine-band comes from
the vertex correction which is relevant if 
$E_{\rm kin} \simle \braket{\omega_q}$, which occurs in the
lightly hole/electron doped region $n\to1$ or $n\to0$.
On the other hand, in the well-doped region we have
$E_{\rm kin} \simge \braket{\omega_q}$ and the vertex corrections
are small.
In the region where doped manganites show ferromagnetism,
vertex corrections do not seem to be important. 
This explains the consistency between experiments and $1/S$ results

\section{COMPARISON WITH EXPERIMENTS}


Abbreviations: Hereafter we use these abbreviations:\\
La$_{1-x}$Sr$_x$MnO$_3$ (LSMO),
La$_{1-x}$Pb$_x$MnO$_3$ (LPMO),
La$_{1-x}$Ca$_x$MnO$_3$ (LCMO),\\
Pr$_{1-x}$Sr$_x$MnO$_3$ (PSMO),
Nd$_{1-x}$Sr$_x$MnO$_3$ (NSMO),
Pr$_{1-x}$Ca$_x$MnO$_3$ (PCMO).

\subsection{Experimental --varieties of properties in ``CMR manganites''--}
\label{Experimental}

Let us briefly mention the varieties of phenomena in CMR manganites.
For a review of recent experiments, readers are referred to
ref.~\citen{Ramirez97}.
It is emphasized that systematic studies of A-site substitution
is quite important to understand the complex behaviors in manganites.
Extrinsic effects due to grain/domain boundaries are also discussed.

\subsubsection{A-site substitution.}
Recent improvements in 
precise control of the A-site cations substitutions in AMnO$_3$
revealed a complex phase diagram as a function of substitution,
temperature and magnetic field. They exhibit various phases with
magnetic, charge, orbital and lattice orderings.
For example,
phase diagram for  doping ($x$) vs. temperature (T) is well known for
 La$_{1-x}$Ca$_x$MnO$_3$ (LCMO),\cite{Schiffer95}
as well as  La$_{1-x}$Sr$_x$MnO$_3$ (LSMO),\cite{Urushibara95}
Nd$_{1-x}$Sr$_x$MnO$_3$ (NSMO) and Pr$_{1-x}$Ca$_x$MnO$_3$
(PCMO).\cite{Tomioka96}
Effects of A-site substitution is also studied for a fixed 
doping.\cite{Hwang95,Moritomo97,Radaelli97}

Major effects of A-site substitution 
are the  bandwidth control and the carrier number 
control.
It is well understood that 
the ratio of rare-earth (3+) ions and alkaline-earth (2+)
ions determines the nominal values of the carrier number $x$.
At the same time,
change of the average radius of the A-site ions
 $\langle r_{\rm A}\rangle$ by chemical substitutions
gives the ``bandwidth control'' through the chemical pressure.
Such kind of chemical control creates
a large change in the nature of the compounds.
In general, compounds with larger $\langle r_{\rm A}\rangle$ 
have higher $T_{\rm c}$.\cite{Hwang95,Moritomo97,Radaelli97}
It is considered to be due to wider
effective bandwidth for $e_{\rm g}$ electrons 
in larger $\langle r_{\rm A}\rangle$ compounds, since
it gives less Mn-O octahedra tilting.
 
However, we should note that
it is still controversial  whether the phase diagram is 
controlled mostly by the bandwidth alone.
For example,
ionic size variation $\sigma( r_{\rm A})$
also plays some role to change $T_{\rm c}$,\cite{Rodriguez-Martinez96}
as well as the fact that 
decrease of $T_{\rm c}$ for small $\langle r_{\rm A}\rangle$
is substantially larger than the estimated
value from the change of bond angles.

\begin{figure}
\epsfxsize=14cm\epsfbox{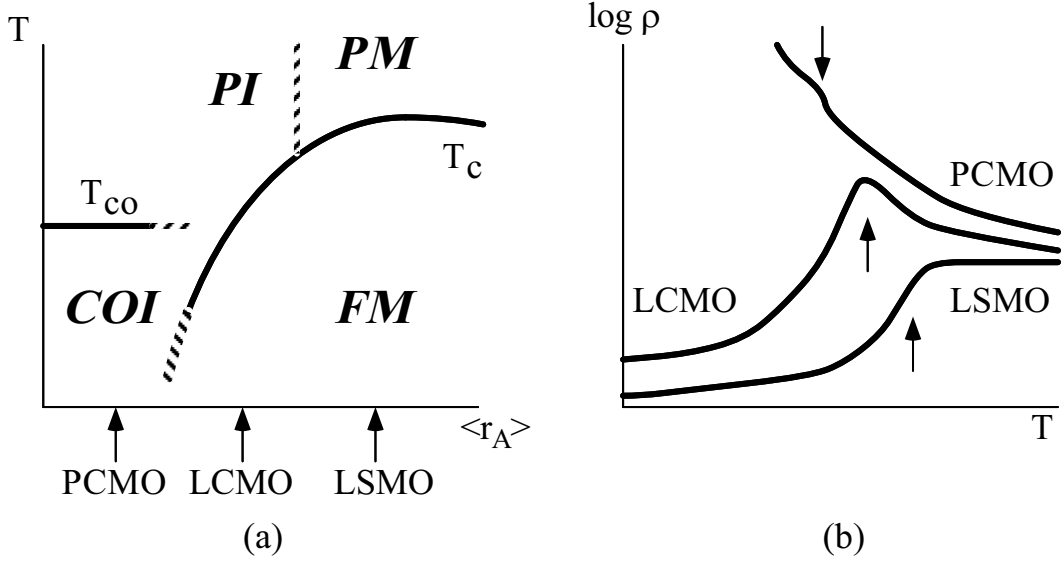}
\caption{(a) Schematic phase diagram at $x\sim 1/3$ by ionic radius
of A-site $\braket{r_{\rm A}}$. Abbreviations are:
paramagnetic metal (PM), paramagnetic insulator (PI), ferromagnetic metal
(FM), charge ordered insulator (COI), as well as 
Curie temperature ($T_{\rm c}$) and charge ordering temperature
($T_{\rm CO}$).
(b) Qualitative behaviors
in $\rho(T)$ for (La,Sr)MnO$_3$ (LSMO), (La,Ca)MnO$_3$ (LCMO)
and  (Pr,Ca)MnO$_3$ (PCMO) at $x\sim 1/3$.
Uparrows ($\uparrow$) in the figure show $T_{\rm c}$,
while downarrows ($\downarrow$) indicates $T_{\rm CO}$.}

\label{FigPhaseD}
\end{figure}

Let us concentrate on the region $x\sim 1/3$ where
it is far from antiferromagnetic insulating phase at $x\sim 0$
and the region with charge and orbital ordering at $x\sim 0.5$.
The compounds are roughly classified as follows:
\begin{itemize}
\item High $T_{\rm c}$ compounds: {\em e.g.} LSMO.\\
A canonical example for the high $T_{\rm c}$ compounds is
(La,Sr)MnO$_3$ (LSMO) with $T_{\rm c} \sim 380{\rm K}$.
Resistivity shows a small value at lowest temperature
($\rho_0 \sim 10^2\mu\Omega\mbox{cm}$).
At $T\sim T_{\rm c}$, $\rho(T)$ takes much larger value
but still in the order of Mott's limit $\rho = 2\sim 4 \mbox{m$\Omega$cm}$.
Above $T_{\rm c}$, $\rho(T)$ shows a metallic behavior,
{\em i.e.} ${\rm d}\rho(T)/{\rm d}T >0$.
Namely, this compound is a good metal below $T_{\rm c}$ and become an
 incoherent metal above $T_{\rm c}$
 with the absolute value for $\rho(T)$
being near Mott's limit.\cite{Urushibara95,Cheong9x}

\item Low $T_{\rm c}$ compounds: {\em e.g.} LCMO, PSMO.\\
LCMO is the most well-investigated compound.
Ca substitution creates smaller $\braket{r_{\rm A}}$ and larger 
$\sigma(r_{\rm A})$.
It has lower $T_{\rm c} \sim 280{\em K}$ compared to LSMO, and
shows metal to insulator transition at around 
$T_{\rm c}$.\cite{Schiffer95,Chahara93,Jin94,Snyder96}

\item Compounds with charge ordering instability: {\em e.g.} PCMO.\\
As $\braket{r_{\rm A}}$ is further decreased,
compounds show a charge ordering 
at $T\sim 200{\rm K}$.\cite{Tomioka95,Tomioka96,Yoshizawa96,Ramirez97}
In the zero-field cooling process, ferromagnetic metal phase 
does not appear.

\end{itemize}

At $x\sim 1/2$, the phase diagram becomes more complex.
At the lowest temperature,
the tendency of the competition between ferromagnetism and
charge ordering\cite{Wollan55,Goodenough55} 
driven by $\braket{r_{\rm A}}$ control
remains the same.
It has been recently discovered that in the intermediate region,
a new phase of A-type antiferromagnetic metal region exists
in the narrow vicinity of $x=1/2$.\cite{Kawano97,Akimoto98}
Temperature dependence is also complex.
Behaviors of the resistivity above $T_{\rm c}$ are
roughly the same with the case of $x\sim 1/3$.
For example, 
metallic behavior above $T_{\rm c}$ is observed for
NSMO at $x=1/2$ as well as compounds with larger $\braket{r_{\rm A}}$.
Substitution
of Nd by Sm reduces $\braket{r_{\rm A}}$ and
makes the paramagnetic phase insulating.\cite{Kuwahara97}

\subsubsection{Extrinsic effects in polycrystal samples.}
In samples with multiple grain structures,
it has been shown that
there exist
so-called tunneling magnetoresistance (TMR) phenomena
through the spin valve
 mechanism.\cite{Snyder96,Hwang96,Gupta96,Raychaudhuri98x} 
Magnetoresistance of the material with artificially
controlled grain/interface
boundaries are also studied to
realize a low-field MR device through TMR.\cite{Sun96,Steenbeck97,Mathur97}

In manganites, the half-metallic behavior due to 
DE interaction  is considered to cause
a large amplitude in MR.
Spin polarizations in grains are schematically
depicted in Fig.~\ref{FigGrain}.
In the zero field case, each grain have different spin orientations.
Namely, the spins of  the half-metallic electrons
are different from one grain to another. Due to the nearly
perfect polarization nature of itinerant electrons,
inter-grain hopping amplitude is suppressed by such random
spin polarizations.
Under the magnetic field, spin polarization of grains become
parallel to the external field. Inter-grain electron hopping becomes
larger in this case. For multi-grain system with half-metallic states
such spin valve phenomena becomes prominent and
gives MR effect in a low field range.

Resistivity in polycrystal samples seems to be dominated
by such extrinsic effects.
One should be careful about discussing the experimental data
from a microscopic point of view.
Such TMR behavior in these perovskite manganites should be
discussed  in relation with
other half-metallic materials such as CrO$_2$ or 
Tl$_2$Mn$_2$O$_7$.\cite{Irkhin94,Khomskii97}

\begin{figure}[htb]
\epsfxsize=8cm\epsfbox{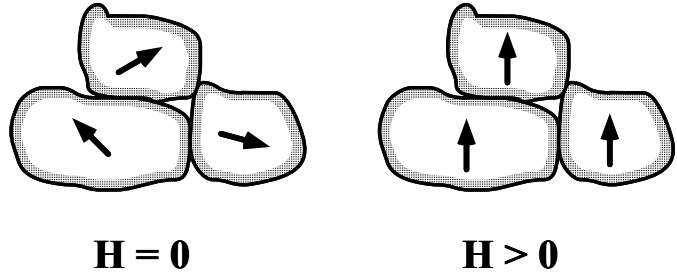}
\caption{Grains of the polycrystal samples. (a) 
at zero magnetic field $H=0$, (b) under magnetic field $H>0$.
Arrows indicate the orientation of the magnetization for
each grains. For the relation with low field TMR, see the text.}
\label{FigGrain}
\end{figure}

\subsection{Comparison with high Curie temperature compounds}
\label{ComparisonHiTc}

Here, we will show the comparison of theoretical results
with experimental data of high $T_{\rm c}$ compounds
such as LSMO and LPMO. We discuss that the DE Hamiltonian
alone explains most of the thermodynamics of these manganites,
including Curie temperature and resistivity.

\subsubsection{Curie temperature.}
In Fig.~\ref{FigTCurieExp} we plot the $x$ dependence of
the  Curie temperature
for LSMO, together with the fitting curves obtained
by DMF.\cite{Furukawa95b} From the fitting, we see that
with parameters $W\sim 1{\rm eV}$ and $\JH/W \sim 4$
the value of $T_{\rm c}$ as well as its $x$-dependence is
reproduced. The bandwidth of $W \sim 1{\rm eV}$ is a
typical value for 3d transition metal oxides, and is consistent
with the band calculation estimate for
 manganites.\cite{Hamada95,Pickett96,Papaconstantopoulos98}
It is also consistent with the value obtained from the
spin wave dispersion fit,\cite{Furukawa96} shown later in this section.

\begin{figure}[htb]
 \epsfxsize=8cm\epsfbox{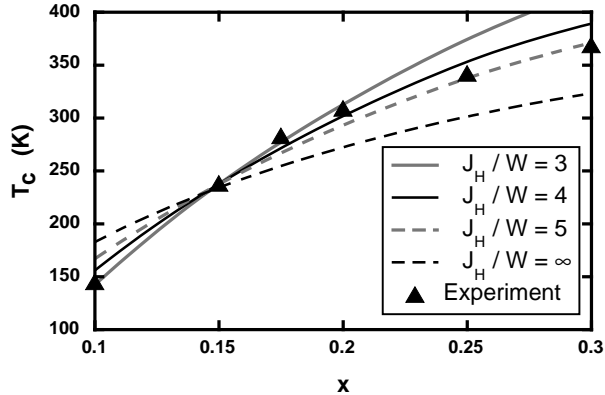}
\caption[]{Curie temperature $T_{\rmc}$ of the DE model
in comparison with those for La$_{1-x}$Sr$_{x}$MnO$_3$.
Experimental data are from ref.~\citen{Tokura94}.
}
\label{FigTCurieExp}
\end{figure}

Thus the value of $T_{\rm c}$ in LSMO is reproduced by the
DE model alone. This result is consistent with other
methods, {\em e.g.} high-temperature expansion
by R{\"o}der {\em et al.}\cite{Roder97} as well as by the
Monte Carlo method.\cite{Yunoki98}

Millis {\em et al.}\cite{Millis95} discussed that
the magnitude of $T_{\rm c}$ as well as its $x$ dependence
for LSMO cannot be accounted for by the DE model alone,
and discussed the importance of the dynamic Jahn-Teller effect.
However, this part of their discussion is due to an inappropriate
estimate of $T_{\rm c}$ which is based on a calculation of
effective spin coupling at $T=0$.
In the itinerant systems, $T_{\rm c}$ is reduced by
spin fluctuation, in general.
Although $T_{\rm c}$ scales with $W$, it is quite small
compared to $W$ due to the prefacter,
whose
typical value is $T_{\rm c} \simle 0.05W$ (See Fig.~\ref{FigTcJHinfty}).
Thus bandwidth of $1{\rm eV}$ creates $T_{\rm c}$ of the order of
room temperature.
Decrease of $T_{\rm c}$ as $x$ decreases is explained by the
reduction of the kinetic energy.\cite{Varma96}

\subsubsection{Resistivity and magnetoresistance.}
Resistivity of the
high-$T_{\rm c}$ compounds in single crystals at sufficiently
large doping $x\sim 1/3$ are different from
those of low-$T_{\rm c}$ compounds or polycrystal samples.
For LSMO at $x\sim 1/3$,\cite{Urushibara95}
 residual resistivity $\rho_0$ is in the order
of a few 10$\mu\Omega{\rm cm}$, and the
temperature dependence $\rho(T)$ shows a monotonously increase,
{\em i.e.} ${\rm d}\rho / {\rm d} T > 0$ even above
$T_{\rm c}$. (For LPMO, see ref.~\citen{Searle69}.
Recent experiment by Cheong {\em et al.}\cite{Cheong9x} shows that
the resistivity of LSMO continuously increase up to 1000K,
without saturation or metal-semiconductor transition.)\ \ 
The value of resistivity at $T_{\rm c}$ is typically
$\rho(T_{\rm c}) = 2\sim 4 {\rm m}\Omega{\rm cm}$,
which is in the order of the Mott limit.
In short, LSMO is a good metal at $T \ll T_{\rm c}$, and
an incoherent metal at $T\simge T_{\rm c}$.
The DE model reproduces these data
 (See Fig.~\ref{FigTempDep}).
Similar temperature dependences of resistivity
are observed in a wide class of materials of half-metals
such as CrO$_2$ and Heusler alloys.\cite{Irkhin94}

It has been discussed that the DE model cannot explain the
resistivity of LSMO in its absolute value as well as the temperature 
dependence.\cite{Millis95} However, it is now clear that
if one compares data for a high quality single crystal
of LSMO (not polycrystal, or other compounds with lower $T_{\rm c}$),
DE alone does account for the resistivity.

\begin{figure}[htb]
\epsfxsize=8cm\epsfbox{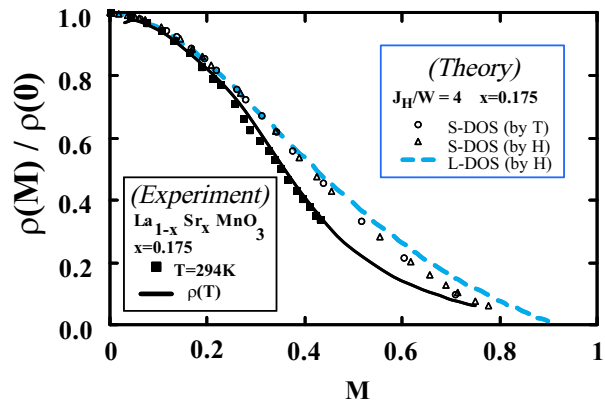}
\caption[]{Magnetoresistance, plotted by 
magnetization vs. resistivity.
Solid curves and filled symbols are the
data for La$_{0.875}$Sr$_{0.175}$MnO$_3$ from ref.~\citen{Tokura94}.
Open symbols and grey curve show DMF results for
$\JH/W=4$ and $x=0.175$, for semicurcular (S-DOS) and Lorentzian (L-DOS).
}

\label{FigMR}
\end{figure}

In Fig.~\ref{FigMR} we show magnetoresistance
of LSMO at $x=0.175$. 
Universal behavior of the magnetoresistance in the form
\begin{equation}
    -\Delta\rho/\rho_0 = C   {M^*}^2
    \label{RhoScaleLSMO}
\end{equation}
is observed, where
$C\sim 4$ 
is temperature/field independent constant.\cite{Urushibara95,Tokura94}
Namely, the resistivity is directly related to the magnetization.
In Fig.~\ref{FigMR} we also show the DMF results.
The curve reproduces the experimental data.\cite{Furukawa95d}
The relation $  {\Delta \rho}/{\rho(0)} = - C M^2$ is 
obtained theoretically in eq.~(\ref{RhoScaleTheory}).
The value $C>1$ shows that the system is in the strong
coupling region.

Note that in  LCMO polycrystal samples, a different scaling relation
\begin{equation}
 \rho(M) = \rho(0) \exp( -\alpha M^2 / T)
  \label{RhoScaleLCMO}
\end{equation}
has been reported.\cite{Hundley95}
From such an activation-type temperature dependence,
polaronic origin of the resistivity has also been discussed.
The difference between the scaling relation in LSMO (\ref{RhoScaleLSMO})
and LCMO (\ref{RhoScaleLCMO}) indicates the qualitative difference
for the origin of CMR.

\subsubsection{Thermoelectric power.}
Seebeck coefficient $S$ for LSMO has been reported.\cite{Asamitsu96}
They show non-universal behaviors, including the change of the sign.
However, in the vicinity of $T_{\rm c}$,
a scaling behavior in the form
$-\Delta S/S(0) \simeq -\Delta\rho/\rho(0)$ irrespective
of doping is reported,
where $\Delta S$ and $\Delta\rho$ are the 
change of Seebeck coefficient and resistivity
under magnetic field, and 
$S(0)$ and $\rho(0)$ are their zero-field values,
respectively.\cite{Asamitsu96}
In Fig.~\ref{FigSeebeck} we plot the DMF result
for the Seebeck coefficient $S$ on a Lorentzian DOS at $\JH/W=4$.
The data is plotted in the form $-\Delta S/S(0)$ vs. $ -\Delta\rho/\rho(0)$.
We see $-\Delta S/S(0) \simeq -\Delta\rho/\rho(0)$ for
different values of $x$.

\begin{figure}
\epsfxsize=8cm\epsfbox{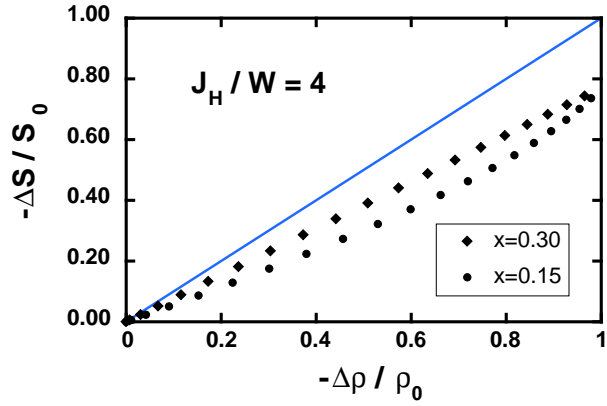}
\caption{Seebeck coefficient under magnetic field,
scaled by resistivity. The line is a guide to eyes.}
\label{FigSeebeck}

\end{figure}

\subsubsection{Spin excitation.}

From the neutron inelastic scattering experiment,
spin wave dispersion relation of La$_{0.7}$Pb$_{0.3}$MnO$_3$ 
(LPMO at $x=0.3$)
is investigated.\cite{Perring96}
LSMO in the ferromagnetic metal 
region has also  been 
studied.\cite{Furukawa98,Martin96,Moudden98,Vasiliu-Doloc97x}

Perring {\em et al.}\cite{Perring96} found that
the experimental
data of spin wave dispersion relation for LPMO fits the
cosine band form. They phenomenologically argued that
a ferromagnetic Heisenberg model
with  nearest-neighbor spin exchange couplings
$ 2 J_{\rm eff} S \sim 8.8{\rm meV}$ 
is a good candidate for the effective 
spin Hamiltonian,
although the material is a metal.

\begin{figure}[htb]

\epsfxsize=8cm\epsfbox{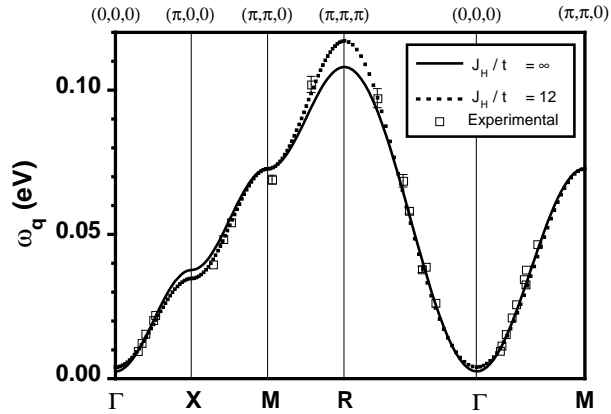}
\caption[]{Spin wave dispersion at $x=0.3$.
The experimental data for La$_{0.7}$Pb$_{0.3}$MnO$_3$ is from
ref.~\citen{Perring96}.}
\label{FigSWdisp}
\end{figure}

Now we discuss the DE model results.\cite{Furukawa96,Kaplan97,Wang98}
In Fig.~\ref{FigSWdisp}, we plot the theoretical results together
with the neutron inelastic scattering experiment data.
As we have discussed previously, cosine-band dispersion
is obtained in the limit $\JH\to\infty$. This gives the
identical fit with the analysis by Perring {\em et al.},
and the fitting parameter determines electron hopping as
 $t \sim 0.3 {\rm eV}$. Here, finite gap at $q=0$ is
artificially introduced as done in ref.~\citen{Perring96},
due to some experimental inaccuracies.

At the zone boundary $q \sim (\pi,\pi,\pi)$ we see the
deviation from the cosine band, or softening of the dispersion.
The data is well accounted for by introducing some finite 
value for $\JH/t$. In Fig.~\ref{FigSWdisp} we show
the result for $\JH/t=12$. From fitting we obtain
$t\sim 0.26 {\rm eV}$ and $\JH \sim 3.1 {\rm eV}$.
Although it is not easy to estimate the
systematic errors of the fitting, it is plausible to say that
the DE model with $t = 0.2\sim 0.3 {\rm eV}$ and 
$\JH \simge 3 {\rm eV}$ explains the spin wave dispersion
relation of LPMO.
These values are the effective hopping energy of the double-exchange
model with reduced degrees of freedom.
Nevertheless, these values of $t$ are consistent with
those estimated from the band 
calculation.\cite{Hamada95,Pickett96,Papaconstantopoulos98}

Anomalous damping of zone boundary magnons are also reported.\cite{Perring96}
Within the DE model, spin wave excitation at finite temperature interacts
with the Stoner continuum as shown in Fig.~\ref{FigImChiZB}.
From eq.~(\ref{ImChiScale}), we see that 
\begin{equation}
  \Gamma(q,T) \propto (1-M^*{}^2) \omega_q,
   \label{MagnonDampScale}
\end{equation}
where $\Gamma(q,T)$ is the linewidth (inverse lifetime) of
the magnon. This explains that the magnons are damped at finite
temperature, especially at the zone boundary.
Indeed such a scaling relation (\ref{MagnonDampScale}) is
observed in LSMO.\cite{Furukawa98}

\subsubsection{Optical conductivity.}
Optical conductivity 
for La$_{0.6}$Sr$_{0.4}$MnO$_3$ by Moritomo {\em et al.}\cite{Moritomo97a}
is shown in Fig.~\ref{FigOcExp}(a).
Here, temperature independent part which is discussed to be due to
$d$-$p$ charge-transfer type excitation is subtracted.
There exists a peak at $\omega \sim 3{\rm eV}$, and its
 temperature dependence is in a way that it vanishes at $T\to 0$.

\begin{figure}[htb]
\par\vspace{-2.5cm}\par
\epsfxsize=12cm\epsfbox{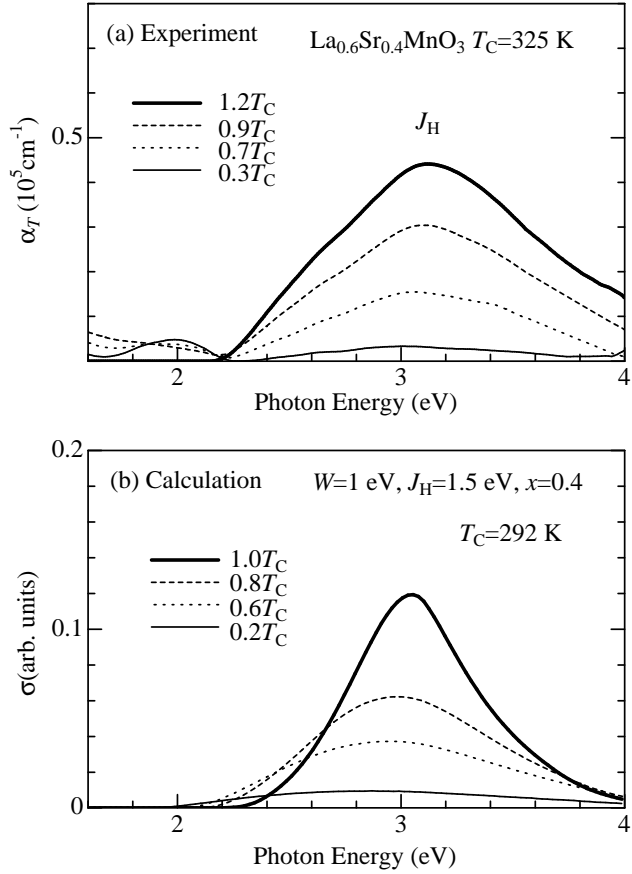}
\par\vspace{-2.5cm}\par
\caption[]{Optical conductivity. (a) Experimental
data for La$_{0.6}$Sr$_{0.4}$MnO$_3$ taken from ref.~\citen{Moritomo97a},
where temperature independent part is subtracted as a background.
(b) DMF result for $W=1{\rm eV}$, $\JH=1.5{\rm eV}$ and $x=0.4$.
}
\label{FigOcExp}
\end{figure}

By choosing $W=1 {\rm eV}$, $\JH=1.5{\rm eV}$ and $x=0.4$,
it is possible to reproduce the experimental data by the DE model.
In Fig.~\ref{FigOcExp}(b), we show the
 DMF results for the semicircular DOS.
The peak structure as well as its temperature dependence is
in agreement.

Let us discuss more details about 
the scaling relation.\cite{Furukawa98x,Moritomo97a}
Experimentally, the integrated spectral weight
\begin{equation}
  S_{\rm exp} \equiv \int _{2.2{\rm eV}}^{4.0{\rm eV}} {\rm d}\omega
   \sigma(\omega)
\end{equation}
shows a scaling relation
\begin{equation}
  S_{\rm exp} \propto 1 - M^*{}^2.
\end{equation}
This is explained by the transfer of DOS
by magnetic fluctuation in the DE model, as shown in eq.~(\ref{OCscaleTheory}).
(See also Fig.~\ref{FigOpt} in \S\ref{DEdynamics}.)

Thus,
optical spectra show experimental evidence for the shift of DOS
which is a typical phenomena in DE model, and give the
estimate of the parameter $\JH\sim 1.5{\rm eV}$.

\subsubsection{Summary: LSMO as a canonical DE system.}
As long as  the high-$T_{\rm c}$ compounds are
concerned, {\em e.g.} LSMO at $x = 0.2 \sim 0.4$,
the DE Hamiltonian accounts for various
experimental data concerning the ferromagnetism and 
the transport
\begin{itemize}
 \item Curie temperature.\cite{Furukawa95b,Yunoki98,Roder97}
 \item Resistivity (absolute value and
       magnetoresistance).\cite{Furukawa94,Furukawa95d}
 \item Spin and charge excitation spectra.\cite{Furukawa95d,Furukawa96}
 \item Scaling behaviors in charge and spin 
       dynamics.\cite{Furukawa98,Furukawa98x}
\end{itemize}
The  mechanism of MR in the single-crystal of these
 high $T_{\rm c}$ compounds are understood from double-exchange alone,
namely due to  spin disorder 
scattering.\cite{Furukawa94,Kubo72,Kasuya56,Fisher68,Nagaev98}

There still exist controversies with some experiments and the DE model.
An example is the loss of spectral weight near the Fermi level 
observed by photoemission experiments,\cite{Park98a,Sarma96}
the other is the optical conductivity
spectrum\cite{Okimoto95} at $\omega \simle 1 {\em eV}$
which shows substantial loss of the Drude weight
(integrated oscilator strength in the low frequency region)
in contradiction with the specific heat behavior which does
not show such a large mass enhancement.
It should be noted that these experiments are highly sensitive to
surface states, and there always remains an open question
whether these experiments really measure the bulk states
or are merely observing the surface states.
Recently, Takenaka {\em et al.}\cite{Takenaka98x} reported that
the optical spectrum of the ``clean'' surface prepared
by cleaving shows larger Drude part compared to those
prepared by surface filing, and is consistently large with
respect to the specific heat estimate.
Although the issue is still far from conclusive results,
we should be very careful about such experimental details.

Let us discuss the value of $\JH$ in manganites estimated from these
calculations.
From experiments concerning spin and charge dynamics,
data fitting gives $\JH = 1.5\sim 2 {\rm eV}$. On the other hand,
static and low frequency
 experiments such as $T_{\rm c}$ and 
$\rho(T)$ measurements are well reproduced by $\JH = 3\sim 5{\rm eV}$.
Such a change in the value of $\JH$ may be understood by
the renormalization of interaction.
While high energy experiments (such as $\sigma(\omega)$)
directly measure the bare value of $\JH$,
low frequency measurement is affected by the dressed quasiparticle
involving the vertex renormalization. In this case, particularly,
the effect of Coulomb interaction should be important.
Since our DE Hamiltonian does not involve on-site Coulomb
repulsion terms, elimination of the double occupancy of
itinerant electrons are underestimated. This is compensated by
increasing the value of $\JH$ which also energetically
prohibits the on-site double occupancy. Thus
the vertex renormalization due to Coulomb repulsion
may increase the value of $\JH$ in the low-frequency scale.

\subsubsection{Discussion: Absence of polaronic behaviors in LSMO.}

Millis {\em et al.}\cite{Millis95} discussed
the discrepancies in DE model and experiments of LSMO
with respect to (i) Curie temperature ($T_{\rm c}$),
(ii) temperature dependence of resistivity ($\rho(T)$),
and (iii) absolute value of resistivity ($\rho(T\sim T_{\rm c})$).
They concluded that DE alone does not explain the MR of LSMO. 

What we have shown here is, however, that
the accurate calculation of the DE model and the
measurements of high quality crystal samples 
indeed give consistent results.
Polaron effects, if any, should be small enough to be irrelevant.
For example, R{\"o}der {\em et al.}\cite{Roder96} discussed
the decrease of $T_{\rm c}$ due to the reduction of
electron hopping caused by the polaronic effect.
Thus, if polaronic interaction is included to the DE model hamiltonian,
$T_{\rm c}$ of the model becomes inconsistently small compared to the
experimental value.

Another point to be discussed is the experimental difference
between LSMO and LCMO (see  also \S\ref{Experimental}).
Since LCMO shows different behaviors in resistivity,
{\em e.g.} temperature dependence of $\rho(T)$ as well as
the MR scaling relations,
different physics must be taking place.
If one assumes the  polaronic arguments,
crossover from metallic (LSMO) to insulating (LCMO) behavior
with respect to the resistivity above $T_{\rm c}$
could be understood in a following way.
Insulating behavior might be due to small-polaron formation
in systems with large electron-lattice coupling,
while systems with smaller electron-lattice coupling
might show large-polaron behaviors.

However, this argument is in contradiction with experimental data.
Pair distribution function (PDF) measurement of atomic 
displacements observed by inelastic neutron scattering\cite{Louca97}
shows that there exist large Jahn-Teller distortions in a dynamic way
even in the metallic phase of LSMO above $T_{\rm c}$.
As long as high $T_{\rm c}$ compound LSMO is concerned,
lattice distortion does not seem to affect 
transport properties as well as magnetic behaviors 
({\em e.g.,} $T_{\rm c}$).
Since large lattice distortion is observed both in
LSMO and LCMO,\cite{Dai96,Billinge96} polaronic
discussion alone is not sufficient to explain the
qualitative difference in $\rho(T)$.

\subsection{CMR effects in low Curie temperature compounds}

What is the origin of the difference in resistivity behavior
of LCMO compared to LSMO?
As shown in \S\ref{ComparisonHiTc}, 
the DE model is sufficient to explain LSMO, and should be
a good reference model to discuss LCMO.
We discuss the half-metallic behavior in ferromagnetic
states and the competition between ferromagnetic metals and
charge ordered insulators.

\subsubsection{Theories of CMR based on the change in transfer integral.}
Several theories based on microscopic models have been proposed
to explain the CMR phenomena in LCMO.
Polaron effects of Jahn-Teller distortions
are introduced to explain the metal-insulator transition
from the point of view of large polaron to small polaron
crossover by magnetism.\cite{Roder96,Millis96l,Millis96b} 
The idea of Anderson localization due to spin 
disorder as well as diagonal charge 
disorder has also been 
discussed.\cite{Varma96,Nagaev98,MullerHartmann96,Allub96,Li97,Sheng97,Kogan98x}

Many of these proposals relate magnetism and transport
through the change of the hopping matrix element 
\begin{equation}
  t_{\rm eff}(T,H) \propto \cos(\theta/2),
\end{equation}
where $\theta$ is the angle between nearest neighbor spins,
as 
discussed by Anderson and Hasegawa.\cite{Anderson55}
Namely, in general, these scenarios discuss the existence of the
critical value of hopping $t_{\rm c}$. 
Change of the mean magnetic structure controls the value of
$t_{\rm eff}$, and it is considered that
some kind of transition occurs at $t_{\rm eff} = t_{\rm c}$.

In a small polaron scenario, there exists a competition between
the electron kinetic energy $E_{\rm kin} \propto t_{\rm eff}$ and
the polaron energy $E_{\rm pol} (\propto \lambda)$.
It is characterized by the critical value for hopping $t_{\rm c}$ where
transition from small polaron state (small $t_{\rm eff}$ region) and
large polaron state (large $t_{\rm eff}$ region) occurs.
Such a change in the behavior of carriers affect the
conductivity through the self trapping mobility or the percolation
of localized carriers.
In the context of Anderson localization,
critical value $t_{\rm c}$ is determined by the
relationship between the mobility edge and the Fermi level.
In either cases,
electron hopping $t_{\rm eff}$ is considered to be controlled by the
magnetic structure through the Anderson-Hasegawa's
DE mechanism.
In the presence of the competition between extended and localized
states,
metallic states are realized  at the large hopping
 region $t_{\rm eff}(T,H) > t_{\rm c}$, and
the system becomes insulating at
 small hopping region $t_{\rm eff}(T,H) < t_{\rm c}$.

However, in order to explain the experiments for resistivity,
these scenario need additional explanations to justify themselves.
Let us point out two issues here:
\begin{itemize}
 \item Does the transition point $t_{\rm c}$  really exist at
       the well-doped system $x=1/3\sim 1/2$?
 \item Why does the ``metal-insulator'' transition occur 
       only in the vicinity of $T_{\rm c}$,
       irrespective of carrier concentration and bandwidth?
\end{itemize}

Localization phenomena in low carrier concentration
is explained by the suppressed overlaps of wavefunctions.
For example, in a low carrier small polaron system,
a self-trapped polaron has a short confinement length scale
compared to mean polaron-polaron distances.
On the other hand, if we consider the region $x\sim1/3$,
every two sites out of total six neighbors in a cubic lattice
is occupied by other polarons. In such a high density system,
quantum mechanical overlaps of polaron wavefunctions 
have to be quite large. In other words,
in order to assume a localized state, the
gain of self-trapping potential energy has to be
unrealistically large to compensate the loss of kinetic
energy by such a localization
with small length scale. For example,
dynamic Jahn-Teller scenario requires that the polaron binding energy
be larger than the electron kinetic energy to make
insulating behaviors for the carrier doped case.\cite{Millis96b}
Thus, it is difficult to understand
 whether the localized state
really exists in a realistic parametrization of the Hamiltonian.

Another point which is hard to explain is the fact that
the metal-insulator transition always occurs in the vicinity of $T_{\rm c}$
for various A-site substitutions
 in both ionic radius and average valence
changes. 
In order to explain the
metal-insulator transition scenarios
require the ``pinning'' of the critical point
at $T\sim T_{\rm c}$, {\em i.e.},
irrespective
of carrier concentration as well as A-site bandwidth and randomness,
 $t_{\rm c} \sim t_{\rm eff}(T=T_{\rm c})$ always has to be satisfied.
Note that $ t_{\rm eff}$ is determined by the short range spin correlation
$\braket{S_i\cdot S_j}$ and is a smooth and continuous function of
temperature without an anomaly at $T_{\rm c}$.
It is also required to explain the
complete absence of the metal-insulator transition in high-$T_{\rm c}$
compounds such as LSMO at $x\sim1/3$ within some realistic
parametrizations.

\subsubsection{Magnetic inhomogeneities
and the nanodomain TMR mechanism.}
In manganites, especially in the low-$T_{\rm c}$ regions,
magnetic inhomogeneities are experimentally observed.
Here we discuss the importance of such inhomogeneities.

Unconventional feature in the low-$T_{\rm c}$ compounds is the
presence of the central peak well below $T_{\rm c}$,\cite{Lynn96}
which indicates the presence of the magnetic cluster and its
diffusive dynamics. From the spin diffusion constant,
the correlation length of the spin clusters are estimated to be
$\xi \sim 10{\rm \AA}$. Neutron elastic scattering
 measurements also observed the
ferromagnetic cluster with correlation length 
$\xi\sim 20{\rm \AA}$.\cite{DeTeresa96,Fernandez-Baca98}

Magnetic inhomogeneities also show up at
the linewidth of the spin wave $\Gamma_q \sim q^2$,\cite{Moudden98}
which systematically increases as $T_{\rm c}$ is suppressed by
A-site substitution.
It is speculated to be due to inhomogeneity effect through
spin stiffness distribution.\cite{Furukawa98,Furukawa98x}
For lower $T_{\rm c}$ compounds, there exist much prominent 
broadening of the spin wave dispersion at the zone boundary.\cite{Hwang98}

Optical conductivity 
measurements\cite{Moritomo97,Kaplan96,Jung98,Quijada98x,Machida98}
show that the A-site substitution causes
changes in the spectrum at the peak structure around $\sim 1.5{\rm eV}$
and the infrared quasi-Drude (incoherent) structures.
Formation of lattice polaron at lower $T_{\rm c}$ compounds
which causes $\omega\sim  1.5{\rm eV}$ peak is discussed.
Inhomogeneities in charge and magnetic
structures observed by $\mu$SR and X-ray 
measurements\cite{Heffner96,Yoon98,Booth98}
suggest that such polaronic cluster remain even at at low temperatures.
Possible micrograin
formation due to charge segregation as well as
phase separation between ferromagnetic and antiferromagnetic domains
has also been discussed.\cite{Allodi97,Hennion98,Perring97}

Thus in most compounds  inhomogeneous behaviors are
observed experimentally. The tendency is that
inhomogeneity is more prominent in the
compounds with lower $T_{\rm c}$ due to
 smaller $\brakets{r_{\rm A}}$ (or larger $\sigma(r_{\rm A})$ as well).

Let us now focus on the case $x\sim 1/3$.
As discussed in \S\ref{Experimental}, decrease of $T_{\rm c}$ is
much larger than the estimate from the reduction of the bandwidth
by $\brakets{r_{\rm A}}$. Inhomogeneity may be playing an important
role to this behavior. Since wide bandwidth compound LSMO
shows a ferromagnetic metal phase while narrow band compound PCMO
is a charge-ordered insulator, there exist a competition
as well as a bistability of these phases in the intermediate
bandwidth region. Inhomogeneous behaviors in LCMO suggests that
coexistence of microscopic domains with ferromagnetic and
charge-ordering correlations might happen.
In the paramagnetic phase, neither correlations are long-ranged
in a macroscopic sense. They should be short ranged and/or dynamic.

In such nanodomain structures with microscopic phase separation,
magnetic phase transition occurs when inter-domain correlations
become long ranged. Although intra-domain correlations may begin
at higher temperature determined by the DE mechanism,
true long range order is controlled by the domain-domain interactions
mediated by the junction structures. Intermediate region with
charge ordering reduces the magnetic coupling between 
ferromagnetic nanodomains. Then $T_{\rm c}$ should be
substantially reduced.
Such nanodomain structure creates two different energy scales
for magnetisms, {\em i.e.} intra-domain and inter-domain interactions.

It also shows up in two correlation lengths. One is the
intra-domain correlation length, which becomes the domain size 
at low temperature regions.
The other is the inter-domain correlation length which
is the length scale to determine the nature of the 
magnetic phase transition.
It has been reported that in the low $T_{\rm c}$ compounds
the magnetic correlation length does not diverge at $T_{\rm c}$
but stays constant, {\em e.g.} $\sim 20{\rm \AA}$
for NSMO.\cite{Fernandez-Baca98}
This unconventional behavior is understood if
the typical length scale of the nanodomain structure
is $\sim 20{\rm \AA}$. Due to the resolution of the
triple-axis experiment, it seems that the macroscopic correlation length
which should diverge at $T_{\rm c}$ was not detected.

Now we discuss the relationship with CMR phenomena.
In the presence of nanodomain structures with ferromagnetic metals
and (charge ordered) insulators, several mechanisms 
create magnetoresistance.
A possible mechanism  is the percolation of metallic nanodomains.
If the external magnetic field is applied, the system
gains energy by increasing  the volume fraction of the ferromagnetic
nanodomains. At the percolation threshold, there exists a
metal-insulator transition.
This scenario is, however, unlikely in the sense that
metal-insulator transition occurs only at the percolation threshold,
while the CMR phenomena is widely observed for various
composition range away from some critical point.

Another possibility to be discussed here is the 
nanodomain TMR phenomena,
as is commonly the case for polycrystals.\cite{Hwang96}
At $T\sim T_{\rm c}$, the
intra-domain spin correlation is well developed so
each nanodomain can behave as a half-metallic domains.
Then, application of magnetic field controls
the spin valve transports between nanodomains.
This scenario is most likely in the sense that
it naturally explains generic MR behavior at $T \sim T_{\rm c}$ and
does not assume any critical points.
The conductivity is controlled by magnetism through
the spin valve channels. Huge sensitivity to the external field
is due to the fact that each metallic nanodomain 
already forms a ferromagnetic cluster.
The idea is consistent with the phenomenological
explanation of resistivity in LCMO
by the Two-fluid model proposed by Jaime and Salomon.\cite{Jaime98x}
They discuss the coexistence of a metallic conductivity path and
an activation-type polaronic conductivity.

Let us finally mention the origin of such nanodomain structures.
Instability of electronic phase separation\cite{Yunoki98}
is a candidate for the initial driving force for such phenomena,
which is stabilized to form droplet structures due to long range
Coulomb interactions.
Another possibility is the effect of static potential disorder
due to A-site cation $R^{3+}$-$A^{2+}$ distributions which
causes charge inhomogeneities,\cite{Varma9x}
 as well as self-trapping effect of lattice polarons.

\section{SUMMARY AND CONCLUDING REMARKS}

In this article we discussed the thermodynamics of the DE model.
We treated the model at finite temperature non-perturbatively
with respect to the spin fluctuations.
Various thermodynamic quantities including
magnetic and transport properties are calculated.
Modifications to the mean-field type treatment
by Anderson-Hasegawa and de Gennes are made.

These new results are found to be very important
in comparing the experimental data for manganites with the DE model.
As long as the high $T_{\rm c}$ compounds ({\em e.g.} LSMO) 
in the metallic phase are concerned, 
the DE model accounts for various experimental 
properties including magnetic transition and resistivity.
Comparison with respect to low $T_{\rm c }$ compounds ({\em e.g.} LCMO)
are also discussed. 

There still remains many open questions.
One is the roles
 of other interactions such as lattice distortion\cite{Millis95}
or orbital fluctuations.\cite{Khomskii97,Ishihara96}
It is experimentally clear that these are quite important
and indeed make some long range orderings
in the insulating phases. However,
as we have shown in this article, such interactions
does not show up in thermodynamic properties in the
high $T_{\rm c}$ metallic phases.
It is quite interesting how and why such ``screening'' of
interactions occur in the metallic phase.

In other words, if we start from a realistic 
model for manganites with orbital and lattice
degrees of freedom as well as Coulomb interactions etc.,
we somehow end up 
with the double-exchange model in the metallic phase
as a consequence of
neglecting high-energy excitations.
Therefore, we should always regard the model as 
the renormalized model with parameters
$ t= t_{\rm eff}$ and $J = J_{\rm eff}$ in eq.~(\ref{HamDXM}).
Or, in a strict sense, we should consider the action
of the double-exchange model in eq.~(\ref{Action}) with
renormalized Green's function $ G_{\rm eff}$
and the coupling strength $J_{\rm eff}$.
Such a detailed renormalization studies will help us
to understand other intermediate phases.

Another point of interest is to understand the nature
of the inhomogeneous systems with nanodomain structures.
Such study with respect to both microscopic and
mesoscopic length scale might help us to understand
the generic features of the physics in 
strongly correlated oxides.

\end{document}